\documentclass[11pt]{article}
\usepackage{geometry}
\usepackage{amsmath}
\usepackage{amssymb}
\usepackage{latexsym}
\usepackage{epsfig}
\usepackage{amsmath,amssymb,amsbsy}
\usepackage{latexsym,layout,amsfonts,amssymb,fontenc,xcolor,amsthm,multirow,amsfonts,bm,textcomp}
\usepackage{hyperref}
\usepackage{accents}
\usepackage{accents}
\usepackage{multirow}
\usepackage{bbding}
\usepackage{graphicx}
\usepackage{pstricks}
\usepackage[utf8x]{inputenc}
\usepackage[english]{babel}
\usepackage{multicol}
\usepackage{cite}
\usepackage{rotating}

\newsavebox{\uuunit}
\sbox{\uuunit}
    {\setlength{\unitlength}{0.825em}
     \begin{picture}(0.6,0.7)
        \thinlines
        \put(0,0){\line(1,0){0.5}}
        \put(0.15,0){\line(0,1){0.7}}
        \put(0.35,0){\line(0,1){0.8}}
       \multiput(0.3,0.8)(-0.04,-0.02){12}{\rule{0.5pt}{0.5pt}}
     \end {picture}}

\def\2{\frac12}
\def\4{\frac14}

\def\equationautorefname~#1\null{eq.~(#1)\null
}
\hypersetup{
    bookmarks=false,             unicode=false,              pdftoolbar=true,            pdfmenubar=true,            pdffitwindow=false,     pdfstartview={FitH},        pdftitle={Non-Supersymmetric Magic Theories and Ehlers Truncations},        pdfauthor={Alessio Marrani, Gianfranco Pradisi, Fabio Riccioni, Luca Romano},         pdfsubject={},                   pdfnewwindow=true,          colorlinks=false,           linkcolor=blue,              citecolor=blue,            filecolor=blue,          urlcolor=blue               linkbordercolor={blue},
    citebordercolor={purple},
    urlbordercolor={gray}
}

\begin{document}

\begin{titlepage}
\begin{center}

\hfill DFPD-2017/TH/01 \\

\vskip 1.5cm

{\Large \bf Non-Supersymmetric Magic Theories \vskip .5cm and Ehlers Truncations}

\vskip 1cm

{\bf Alessio Marrani\,$^{1,2}$, Gianfranco Pradisi\,$^{3}$, Fabio Riccioni\,$^4$, Luca Romano\,$^5$}

\vskip 25pt

{\em $^1$ \hskip -.1truecm Museo Storico della Fisica e Centro Studi e Ricerche “Enrico Fermi” \\
Via Panisperna 89A, I-00184, Roma, Italy
 \vskip 15pt }

 {\em $^2$ \hskip -.1truecm Dipartimento di Fisica e Astronomia, Universit\`a di Padova,\\
 and INFN, Sez. di Padova,\\
Via Marzolo 8, I-35131, Padova, Italy
 \vskip 15pt }

{\em $^3$ \hskip -.1truecm Dipartimento di Fisica, Universit\`a di Roma ``Tor Vergata'',\\
and INFN, Sez. di Roma ``Tor Vergata''\\
Via della Ricerca Scientifica 1, I-00133 Roma, Italy
 \vskip 15pt }

{\em $^4$ \hskip -.1truecm
 INFN Sezione di Roma,  Dipartimento di Fisica, Universit\`a di Roma ``La Sapienza'',\\ Piazzale Aldo Moro 2, I-00185 Roma, Italy
 \vskip 15pt }

{\em $^5$ \hskip -.1truecm
Riemann Center for Geometry and Physics\\
Leibniz Universitaet Hannover\\
Appelstrasse 2, D-30167 Hannover, Germany  \vskip 15pt }

{email addresses: {\tt alessio.marrani@pd.infn.it}, {\tt  Gianfranco.Pradisi@roma2.infn.it},  {\tt Fabio.Riccioni@roma1.infn.it}, {\tt lucaromano2607@gmail.com } \\}

\end{center}

\vskip 0.5cm

\begin{center} {\bf ABSTRACT}\\[3ex]
\end{center}
We consider the non-supersymmetric ``magic'' theories based on the split quaternion and the split complex division algebras. We show that these theories arise as ``Ehlers'' $SL(2,\mathbb{R})$ and $SL(3,\mathbb{R})$ truncations of the maximal supergravity theory, exploiting techniques related to very-extended  Kac-Moody algebras. We also generalise the procedure to other $SL(n,\mathbb{R})$ truncations, resulting in additional classes of non-supersymmetric theories, as well as to truncations of  non-maximal theories. Finally, we discuss duality orbits of extremal black-hole solutions in some of these non-supersymmetric theories.

\end{titlepage}

\newpage \setcounter{page}{1}

\tableofcontents

\newpage

\setcounter{page}{1} \numberwithin{equation}{section}

\section{Introduction}

Supergravity theories possess hidden global symmetries that turn out to be
much larger than those expected from the geometry of the compactification
space. For instance, maximal supergravities in $D\geq 3$ exhibit an $%
E_{11-D(11-D)}$ \cite{Cremmer:1978ds, Cremmer:1979up} global symmetry, to be compared with
the $GL(11-D,\mathbb{R})$ group related to the isometries of the $(11-D)$%
-dimensional torus. Extra symmetries, of course, are not necessarily
symmetries of the full Lagrangian, but they leave the field equations
invariant. In relation to string theory, it is conjectured \cite{Hull:1994ys}
that the full non-perturbative string models are invariant only under
discrete subgroups of the global symmetries of the corresponding low-energy
supergravity theories. Such discrete symmetries are usually termed
U-dualities, and they play a crucial role in understanding the deep
relations between different perturbative string theories, that are mapped
under their action one to the other in diverse regions of the moduli space
\cite{Schwarz:1996bh}. Hidden global symmetries are also useful in determining
all possible massive deformations of a given ungauged supergravity. Indeed,
the embedding tensor \cite{Nicolai:2000sc,deWit:2002vt,deWit:2004nw,deWit:2005ub,Samtleben:2005bp} singles out the allowed gauge
groups inside the global symmetry group, providing a covariant description
of all possible gaugings of a given supergravity theory.

\begin{figure}[h]
\centering
\scalebox{0.35} 
{\
\begin{pspicture}(0,-1.14375)(39.1875,4.10375)
\psline[linewidth=0.02cm](1.926875,-0.29625)(38.126877,-0.29625)
\psline[linewidth=0.02cm](30.126875,3.70375)(30.126875,-0.29625)
\pscircle[linewidth=0.02,dimen=outer,fillstyle=solid](2.126875,-0.29625){0.4}
\pscircle[linewidth=0.02,dimen=outer,fillstyle=solid](6.126875,-0.29625){0.4}
\pscircle[linewidth=0.02,dimen=outer,fillstyle=solid](10.126875,-0.29625){0.4}
\pscircle[linewidth=0.02,dimen=outer,fillstyle=solid](14.126875,-0.29625){0.4}
\pscircle[linewidth=0.02,dimen=outer,fillstyle=solid](18.126875,-0.29625){0.4}
\pscircle[linewidth=0.02,dimen=outer,fillstyle=solid](30.126875,3.70375){0.4}
\pscircle[linewidth=0.02,dimen=outer,fillstyle=solid](26.126875,-0.29625){0.4}
\pscircle[linewidth=0.02,dimen=outer,fillstyle=solid](22.126875,-0.29625){0.4}
\rput(2.47875,-1.28625){\huge 1}
\rput(26.490938,-1.28625){\huge 7}
\rput(22.503124,-1.28625){\huge 6}
\rput(18.491875,-1.28625){\huge 5}
\rput(14.512813,-1.28625){\huge 4}
\rput(10.4948435,-1.28625){\huge 3}
\rput(6.4889064,-1.28625){\huge 2}
\rput(38.69875,-1.28625){\huge 10}
\pscircle[linewidth=0.02,dimen=outer,fillstyle=solid](30.126875,-0.29625){0.4}
\pscircle[linewidth=0.02,dimen=outer,fillstyle=solid](38.126877,-0.29625){0.4}
\pscircle[linewidth=0.02,dimen=outer,fillstyle=solid](34.126877,-0.29625){0.4}
\rput(30.501875,-1.28625){\huge 8}
\rput(34.707813,-1.28625){\huge 9}
\rput(30.702969,2.91375){\huge 11}
\pscircle[linewidth=0.02,dimen=outer,fillstyle=solid](10.126875,-0.29625){0.4}
\pscircle[linewidth=0.02,dimen=outer,fillstyle=solid](6.126875,-0.29625){0.4}
\end{pspicture}
}
\caption{{\protect\footnotesize The $E_{8(8)}^{+++}$ Dynkin diagram.}}
\label{E88+++dynkindiagram}
\end{figure}
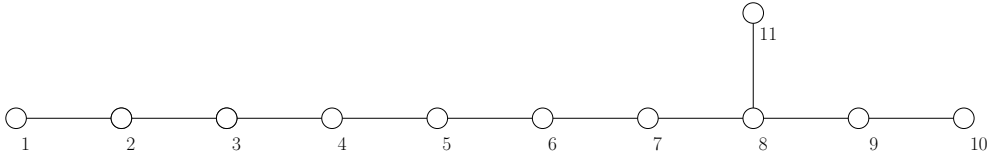

The maximal supergravity theories in any dimension are related to the
very-extended Kac-Moody algebra $E_{8(8)}^{+++}$ (also called $E_{11}$) \cite{West:2001as}, whose Dynkin diagram is reported in fig. \ref{E88+++dynkindiagram}.
The maximal theory in dimension $D$ corresponds to the decomposition of the
algebra in which the ``gravity line'' is
identified with the $A_{D-1}$ subalgebra containing node 1, while the part
of the diagram that is not connected to the $A_{D-1}$ subalgebra corresponds
to the internal symmetry. From the diagram one then sees that the highest
dimension to which the theory can be uplifted is 11, while there are two
different theories in ten dimensions, namely the IIA theory (with $A_{9}$
given by nodes from 1 to 9) and the IIB theory (with $A_{9}$ given by nodes
from 1 to 8 plus node 11) \cite{Schnakenburg:2001he}. Decomposing the
adjoint representation of $E_{8(8)}^{+++}$ in any dimension one obtains all
the $p$-forms of the theory. This not only includes all the propagating
fields and their duals, but also $(D-1)$-forms and $D$-forms. The $(D-1)$-forms
are dual to the massive/gauge deformations, while the $D$-forms are related
to space-filling branes. In particular, in the IIA case one obtains an 8-form
which is dual to the Romans mass \cite{Schnakenburg:2002xx}, and
generalising this analysis to any dimension one discovers \cite%
{Riccioni:2007au,Bergshoeff:2007qi} that the $(D-1)$-forms predicted in $D$
dimensions by $E_{8(8)}^{+++}$ precisely correspond to the representations
of the embedding tensor derived in \cite{Nicolai:2000sc, deWit:2002vt, deWit:2004nw,deWit:2005ub, Samtleben:2005bp}. All the $p$-forms
with the corresponding representations for the maximal supergravities are
reported in Table \ref{E8+++spectrum} (taken from \cite%
{Riccioni:2007au,Bergshoeff:2007qi}).
\begin{table}[h]
\renewcommand{\arraystretch}{1.1}
\par
\begin{center}
\scalebox{.75}{
\begin{tabular}{|c|c||c|c|c|c|c|c|c|c|c|c|}
\hline
Dim & Symmetry  & $p=1$  & $p=2$ & $p=3$ & $p=4$ & $p=5$ & $p=6$ & $p=7$ & $p=8$ & $p=9$ & $p=10$ \\
\hline
\hline
11 & $-$ &  &   &  ${\bf 1}$  &  &  & ${\bf 1}$ &  & & & \\
\hline
10A & $\mathbb{R}^+$ & ${\bf 1}$ & ${\bf 1}$ & ${\bf 1}$ & & ${\bf 1}$ & ${\bf 1}$  & ${\bf 1}$ & ${\bf 1}$ & ${\bf 1}$ & $2\times {\bf 1}$\\
\hline
\multirow{2}{*}{ 10B }& \multirow{2}{*}{$SL(2,\mathbb{R})$} &  &\multirow{2}{*}{ ${\bf 2}$} &  & \multirow{2}{*}{${\bf 1}$}  & & \multirow{2}{*}{${\bf 2}$}  & & \multirow{2}{*}{${\bf 3}$}  &  & $ {\bf 4}$\\
 &  &  &  &  & & &   & &   &  & $ {\bf 2}$\\
\hline \multirow{2}{*}{$9$} & \multirow{2}{*}{$GL(2,\mathbb{R})$} & ${\bf 2}$ & \multirow{2}{*}{${\bf 2}$} & \multirow{2}{*}{${\bf 1}$} &  \multirow{2}{*}{${\bf 1}$}  & \multirow{2}{*}{${\bf 2}$} & ${\bf 2}$ &${\bf 3}$ & ${\bf 3}$ & ${\bf 4}$
 \\
 & & ${\bf 1}$& & &&  & ${\bf 1}$ & ${\bf 1}$& ${\bf 2}$& $2 \times {\bf 2}$ \\
 \cline{1-11}
\multirow{3}{*}{$8$} & \multirow{3}{*}{$SL(3,\mathbb{R})\times SL(2,\mathbb{R})$ } & \multirow{3}{*}{${\bf (\overline{3},2)}$} & \multirow{3}{*}{${\bf ({3},1)}$}  & \multirow{3}{*}{${\bf ({1},2)}$}  & \multirow{3}{*}{${\bf (\overline{3},1)}$}  & \multirow{3}{*}{${\bf ({3},2)}$} & ${\bf (8,1)}$& ${\bf (6,2)}$ & ${\bf (15,1)}$\\
& & & & & & &  & & ${\bf (3,3)}$\\
&&&&&&& ${\bf ({1},3)}$& ${\bf (\overline{3},2)}$& $2\times {\bf (3,1)}$\\
\cline{1-10}
\multirow{3}{*}{$7$} & \multirow{3}{*}{$SL(5,\mathbb{R})$} &\multirow{3}{*}{${\bf \overline{10}}$} & \multirow{3}{*}{${\bf 5}$} & \multirow{3}{*}{${\bf \overline{5}}$} & \multirow{3}{*}{${\bf 10}$} & \multirow{3}{*}{${\bf 24}$} & ${\bf \overline{40}}$&  $ {\bf 70}$  \\
& & &  & &  &  & & ${\bf {45}}$  \\
& & & & &  &  & ${\bf \overline{15}}$& ${\bf {5}}$  \\
\cline{1-9}
\multirow{3}{*}{$6$} & \multirow{3}{*}{$SO(5,5)$} &\multirow{3}{*}{${\bf 16}$} & \multirow{3}{*}{${\bf 10}$} & \multirow{3}{*}{${\bf \overline{16}}$} & \multirow{3}{*}{${\bf 45}$} &\multirow{3}{*}{${\bf 144}$} & $ {\bf 320}$ \\
& & &  & &  & & ${\bf \overline{126}}$ \\
& & & & &  &  & ${\bf {10}}$ \\
\cline{1-8}
\multirow{2}{*}{$5$} & \multirow{2}{*}{$E_{6(6)}$} &\multirow{2}{*}{${\bf 27}$} & \multirow{2}{*}{${\bf \overline{27}}$} & \multirow{2}{*}{${\bf 78}$} & \multirow{2}{*}{${\bf 351}$} &${\bf  \overline{1728}}$  \\
 & & &  & &  &${\bf \overline{27}}$  \\
\cline{1-7}
\multirow{2}{*}{$4$} & \multirow{2}{*}{$E_{7(7)}$} &\multirow{2}{*}{${\bf 56}$} & \multirow{2}{*}{${\bf 133}$} & \multirow{2}{*}{${\bf 912}$} & ${\bf 8645}$ \\
 & & &  & & ${\bf {133}}$ \\
\cline{1-6}
\multirow{3}{*}{$3$} & \multirow{3}{*}{$E_{8(8)}$} &\multirow{3}{*}{${\bf 248}$} & ${\bf 3875}$ & ${\bf 147250}$  \\
 & & & & ${\bf 3875}$ \\
& & & ${\bf 1}$& ${\bf 248}$ \\
\cline{1-5}
\end{tabular}
}
\end{center}
\caption{{\protect\footnotesize All the $p$-forms of the }$E_{8(8)}^{+++}$
{\protect\footnotesize theory in any dimension.}}
\label{E8+++spectrum}
\end{table}

The same construction can be extended to $1/2$-maximal and symmetric $1/4$%
-maximal supergravities where, precisely as in the maximal theory, the
Kac-Moody algebra is $G_{3}^{+++}$, $G_{3}$ being the global symmetry of
the three-dimensional theory \cite%
{Schnakenburg:2004vd,Bergshoeff:2007vb,Riccioni:2008jz}. In particular, the $%
1/2$-maximal theories correspond to the Kac-Moody algebra $SO(8,n)^{+++}$,
while among the $1/4$-maximal theories one can consider the magic ones \cite%
{Gunaydin:1983rk,Gunaydin:1983bi,Gunaydin:1984ak,Bellucci:2013kda},
corresponding to the algebras $F_{4(4)}^{+++}$, $E_{6(2)}^{+++}$, $%
E_{7(-5)}^{+++}$ and $E_{8(-24)}^{+++}$ \cite{Riccioni:2008jz}. With the
exception of the $F_{4(4)}^{+++}$ case, the real form of the symmetry group
of these theories is non-split, a fact taken into account by
considering the Tits-Satake diagram of $G_{3}^{+++}$, which identifies the
real form of the global symmetry in any dimension, as well as the highest
dimension to which the theory can be uplifted \cite{Riccioni:2008jz}.

As shown in \cite{Gunaydin:1983rk,Gunaydin:1983bi,Gunaydin:1984ak,Bellucci:2013kda}, the
exceptional $\mathcal{N}=2$ Maxwell-Einstein supergravities in $D=4,5$ are
related to cubic Jordan algebras. In particular, they are based on simple
Euclidean Jordan algebras $\mathbf{J}_{3}^{\mathbb{A}}$ generated by $3\times 3$ Hermitian matrices over the four normed division algebras $\mathbb{A}=\mathbb{R},\mathbb{C},\mathbb{H},\mathbb{O}$. They can be
associated to the \textit{single split} version of the famous
``magic square'' of Freudenthal, Rozenfeld
and Tits \cite{FRT}. For this reason they are called ``magic''. The complex and quaternionic theories are
consistent $\mathcal{N}=2$ truncations of the maximal $\mathcal{N}=8$
supergravity, while the $\mathcal{N}=2$ octonionic theory is not, as
manifested by the fact that in four dimensions it is based on the minimally
non-compact real form $E_{7(-25)}$, completely different from the maximally
non-compact (split) real form $E_{7(7)}$.

An analogous analysis can be performed by replacing the division algebras
with their \textit{split} versions ${\mathbb{A}}_{s}$ and the magic square
with the \textit{doubly split} magic square \cite{Barton:2000ki} (also
\textit{cfr.} \cite{Cacciatori:2012cb}) given in Table \ref{doublysplitmagicsquare}.
\begin{table}[t!]
\renewcommand{\arraystretch}{1.1}
\par
\begin{center}
\begin{tabular}{|c||c|c|c|c|}
\hline
 & $\mathbb{R}$  & $\mathbb{C}_s$  & ${\mathbb{H}}_{s}$ & ${\mathbb{O}}_{s}$   \\
\hline
\hline
$\mathbb{R}$ & $SO(3)$ & $SL(3,\mathbb{R})$ & $Sp(6,\mathbb{R})$ & $F_{4(4)}$\\
\hline
$\mathbb{C}_s$  & $SL(3,\mathbb{R})$ & $SL(3,\mathbb{R}) \times SL(3,\mathbb{R})$ & $SL(6,\mathbb{R})$ & $E_{6(6)}$ \\
\hline
${\mathbb{H}}_{s}$ & $Sp(6,\mathbb{R})$ & $SL(6,\mathbb{R})$ & $SO(6,6)$ & $E_{7(7)}$ \\
\hline
${\mathbb{O}}_{s}$  &  $F_{4(4)}$ & $E_{6(6)}$  & $E_{7(7)}$ & $E_{8(8)}$\\
\hline
\end{tabular}
\end{center}
\caption{{\protect\footnotesize The doubly split magic square \cite{Barton:2000ki}.}}
\label{doublysplitmagicsquare}
\end{table}
 While the theory based over the split octonions $\mathbb{O}_{s}$ is the maximal supergravity theory, the theories based on ${\mathbb{C}}_{s}$ and ${\mathbb{H}}_{s}$ are non-supersymmetric and their field content
is a gravitational model not interpretable as the bosonic sector of a
locally supersymmetric theory. Exactly as for all the supersymmetric
theories discussed above, they correspond to the very-extended
Kac-Moody algebras $E_{6(6)}^{+++}$ and $E_{7(7)}^{+++}$. These
algebras were originally considered in \cite{Englert:2003zs} and, as emerging from their 
diagram, they can be uplifted at most to eight and ten
dimensions, respectively. The spectrum of forms of the $E_{6(6)}^{+++}$
theory in eight dimensions, as well as the one of the $E_{7(7)}^{+++}$
theory in nine and ten dimensions, was listed in \cite{Kleinschmidt:2003mf}.
In the same paper it was also shown that these Kac-Moody algebras are
consistent truncations of $E_{8(8)}^{+++}$.

It is the main purpose of this paper to further investigate these
truncations. In particular, we refer to the analysis of \cite{Ferrara:2012zc}, where it is observed that in any supersymmetric theory with scalars
parametrising a symmetric manifold, the global symmetry group $G_{3}$ in
three dimensions can be decomposed in $D$ dimensions factorizing the Ehlers
group $SL(D-2,\mathbb{R})$ as
\begin{equation}
G_{3}\supset G_{D}\times SL(D-2,\mathbb{R})\ .  \label{ehlers}
\end{equation} 
The group on the right-hand side of \eqref{ehlers} was dubbed ``super-Ehlers
group'' in \cite{Ferrara:2012zc}. In particular, for the maximal theories,
one obtains the so-called Cremmer-Julia sequence (\textit{cfr. e.g.} Sec. 1
of \cite{Marrani:2010de}) 
\begin{equation}
E_{8(8)}\supset E_{11-D(11-D)}\times SL(D-2,\mathbb{R})\ .  \label{N=8ehlers}
\end{equation}
Within the sequence in eq. \eqref{N=8ehlers}, the  $E_{7(7)}$ and $E_{6(6)}$ groups that appear  in the fourth column of
the doubly split magic square in Table \ref{doublysplitmagicsquare} occur as  
U-duality groups of the maximal supergravity theory in $D=4$ and $D=5$ respectively.  Notice that the same groups also occur in the fourth row of the magic square; thus, as mentioned
before, the theories based on ${\mathbb{H}}_{s}$ and ${\mathbb{C}}_{s}$, that can be obtained by the very-extended Kac-Moody
algebras $E_{7(7)}^{+++}$ and $E_{6(6)}^{+++}$, have in three dimensions
precisely the symmetry of the maximal theory in four and five dimensions, 
respectively. The truncation of the $E_{8(8)}^{+++}$ theory that leads to
the $E_{7(7)}^{+++}$ and $E_{6(6)}^{+++}$ ones is therefore precisely of
Ehlers type. Thus, it is natural to compare the $p$-form spectra of the $E_{7(7)}^{+++}$
and $E_{6(6)}^{+++}$ algebras in any dimension with those obtained 
by truncating in any dimensions the representations of the maximal theories (given in Table \ref{E8+++spectrum}) to singlets of $SL(2,\mathbb{R})$ and $SL(3,\mathbb{R})$. We obtain a perfect match, with the only exceptions of the
multiplicity of the lower-dimensional representations of the $D$-forms in $D$
dimensions and of the 2-forms in three dimensions, which are typically less
than the result of the truncation. We use the software \texttt{SimpLie} \cite{Bergshoeff:2007qi} to determine the spectrum of the theories in any
dimension.

Our procedure can be extended to Ehlers $SL(n,\mathbb{R})$ truncations with
$n>3$, giving again theories that in three dimensions have the symmetry of
the maximal theory in $n+2$ dimensions. As a consequence, their spectra are 
given by the Kac-Moody algebras $SO(5,5)^{+++}$ ($n=4$), $SL(5,\mathbb{R})^{+++}$ ($n=5$) and $(SL(3,\mathbb{R}) \times SL(2,\mathbb{R}) )^{+++}$ ($n=6$).  While the first two were discussed in \cite{Kleinschmidt:2003mf}, the third is the extension of a group which is not simple. The general
analysis of Kac-Moody algebras of this type was first considered in \cite{Kleinschmidt:2008jj}. We compare the spectrum of these theories with the
Ehlers truncation of the spectrum of the maximal theory in any dimensions, obtaining the same match as for the $n=2$ and $n=3$ cases.  In particular, one recovers the ``split magic triangle'' of \cite{Cremmer:1999du}.

An allowed realization of these models within perturbative string theory is possible only if the Ehlers $SL(n,\mathbb{R})$ truncation can be embedded in
the perturbative symmetry of the maximal theory, \textit{i.e.} the one that
does not act on the string dilaton. We show how this can be directly
extracted by an analysis of the Dynkin diagram of the very-extended algebra of the truncated theory. The fact that such Dynkin diagrams encode information about T-duality transformations was originally shown in \cite{Kleinschmidt:2003mf}.

The truncation procedure can also be applied in the same fashion to
1/2-maximal and to 1/4-maximal theories. In \cite{Breitenlohner:1987dg} a large class of symmetric $G/H$ non-linear $\sigma $%
-models has been analysed in $D=3$ and lifted to $D=4$. The cosets are
non-compact Riemannian symmetric spaces, and some of them correspond to the
bosonic sectors of supergravities with sixteen or eight supersymmetries. They
can be extended to the possible allowed higher dimensions using again the
technique based on the related very-extended Kac-Moody algebras.

For the theories based on ${\mathbb{H}}_{s}$ and ${\mathbb{C}}_{s}$, one can
use the connection with Jordan algebras to determine the orbits of extremal
black hole solutions in four and five dimensions. In \cite{Lu:1997bg} the analysis of black-hole orbits of the maximal theory was
performed in terms of bound states of the weights of the representations to
which the black-hole charges belong. By repeating the same analysis for the representations of the black hole charges of the $E_{6(6)}^{+++}$ and $E_{7(7)}^{+++}$ theories, one obtains not only a perfect match with the Jordan algebra investigation, but also an extension of it to higher
dimensions. Following \cite{Marrani:2015gfa}, where it was shown how the results of \cite{Lu:1997bg} can be extended to non-split groups, one can also determine the orbits of the truncations of the 1/2-maximal and the 1/4-maximal theories.

The paper is organised as follows.
In Section \ref{nonsusytheories} we analyse the spectrum of the theories
based on the very-extended Kac-Moody algebras $E_{7(7)}^{+++}$ and $E_{6(6)}^{+++}$. We prove that they are consistent truncations of maximal
theories (based on $E_{8(8)}^{+++}$) obtained by modding out the
corresponding Ehlers group. We also discuss the embeddings within string
theory. In Section \ref{Maximaltrunc}, we extend the class of theories by
observing that the duality group in $D=3$ exactly coincides with that of
the Ehlers truncation of the maximal theory in dimension\footnote{As mentioned, at least where exceptional Lie algebras are involved, this is
a consequence of the symmetry of the doubly split magic square \cite{Barton:2000ki}.} $D$. In the explained sense, our procedure \textit{gives an alternative definition}
of the very-extended duality algebra. When considering non-simple duality
algebras, our results coincide with those obtained using the method proposed
in \cite{Kleinschmidt:2008jj}. In Section \ref{nonmaximaltrunc}, we extend
the construction to a large class of theories with less supersymmetry, like
the ones discussed in \cite{Breitenlohner:1987dg}, and to the infinite
series of orthogonal very-extended algebras, related also to the dimensional
reduction of the heterotic string. In Section \ref{BHorbits}, the analysis
of \cite{Marrani:2015gfa} is extended to the black-hole orbits of the
models presented in this paper. As obtained in \cite{Marrani:2015gfa}, the absence of orbit splitting can be traced back to the lack of short
weights in the corresponding representation space. Finally, in Section \ref{conclusions} we discuss the embedding within string theory, and we present
our conclusions and discussions about possible future directions.

\section{Non-supersymmetric theories based on $\mathbb{C}_s$ and $\mathbb{H}%
_s$}\label{nonsusytheories}

The $\mathcal{N}=2$ ``magic''
Maxwell-Einstein supergravity theories discovered in \cite{Gunaydin:1983rk,Gunaydin:1983bi,Gunaydin:1984ak} are based on the simple
Euclidean Jordan algebras $\mathbf{J}_{3}^{\mathbb{A}}$ generated by $3\times 3$ Hermitian matrices over the four normed division algebras $\mathbb{A}=\mathbb{R},\mathbb{C},\mathbb{H},\mathbb{O}$. The symmetry of the
theories in dimension from three to five is connected to the single spit
form of the famous ``magic square'' of
Freudenthal, Rozenfeld and Tits (\textit{cfr.} \cite{FRT}, as well as the
recent review on magic squares of order 3 and their relevance in
(super)gravity theories \cite{Cacciatori:2012cb}). In particular, in three
dimensions these theories have global symmetry $F_{4(4)}$, $E_{6(2)}$, $E_{7(-5)}$ and $E_{8(-24)}$, respectively. The relation between these
theories and the Jordan algebras $\mathbf{J}_{3}^{\mathbb{A}}$ was used in
\cite{Ferrara:1997uz} (see also \cite{Borsten:2011ai}) to classify the orbits
of all the extremal black-hole solutions in various dimensions. An analogous
magic square exists also for the split division algebras $\mathbb{R},\mathbb{C}_{s},\mathbb{H}_{s},\mathbb{O}_{s}$ \cite{Barton:2000ki}: with complex
parameters, the corresponding groups are the same as those resulting from
the division algebras, but they enter now in the split real form. The
split-octonion case corresponds to the $E_{8(8)}$ symmetry in three
dimensions and therefore the algebra is that associated to maximal
supergravity. In this section, we want to consider the remaining
theories based on $\mathbb{C}_{s}$ and $\mathbb{H}_{s}$; in three
dimensions, they have a (electric-magnetic) duality\footnote{Here duality is referred to as the analogue in a non-supersymmetric
context of the ``continuous'' symmetries
of \cite{Cremmer:1978ds, Cremmer:1979up}, whose discrete versions in the supersymmetric case are the U-dualities of non-perturbative string theory introduced by Hull and Townsend \cite{Hull:1994ys}.} symmetry groups $E_{6(6)}$ and $E_{7(7)}$, respectively.

In \cite{Riccioni:2008jz} it was shown that the bosonic spectrum of the 
$\mathcal{N}=2$ magic theories can be derived from the Kac-Moody algebras $G^{+++}$ where $G$ is the symmetry of the three-dimensional theory. The real
form of the symmetry group in any dimension is identified by the
corresponding Tits-Satake diagram, and the symmetry of the theory in a given
dimension is obtained by the deletion of the corresponding node precisely as
in the maximal case. The deleted node is in general associated to the global scale symmetry and therefore it must correspond to a non-compact Cartan generator, and while for split real forms one can always take the Cartan generators to be all non-compact, for other real forms this is not the case and the Tits-Satake diagram precisely identifies which Cartan generators are compact and which are non-compact. For the Kac-Moody algebras associated to the 
$\mathcal{N}=2$ magic theories, the 
fact that  the nodes of the Tits-Satake diagrams corresponding to compact Cartan generators  cannot be deleted explains from this perspective why these theories can only be uplifted at most to six
dimensions.

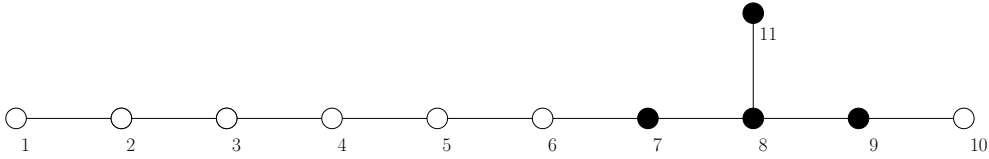
\begin{figure}[h]
\centering
\scalebox{0.35} 
{\
\begin{pspicture}(0,-1.14375)(39.1875,4.10375)
\psline[linewidth=0.02cm](1.926875,-0.29625)(38.126877,-0.29625)
\psline[linewidth=0.02cm](30.126875,3.70375)(30.126875,-0.29625)
\pscircle[linewidth=0.02,dimen=outer,fillstyle=solid](2.126875,-0.29625){0.4}
\pscircle[linewidth=0.02,dimen=outer,fillstyle=solid](6.126875,-0.29625){0.4}
\pscircle[linewidth=0.02,dimen=outer,fillstyle=solid](10.126875,-0.29625){0.4}
\pscircle[linewidth=0.02,dimen=outer,fillstyle=solid](14.126875,-0.29625){0.4}
\pscircle[linewidth=0.02,dimen=outer,fillstyle=solid](18.126875,-0.29625){0.4}
\pscircle[linewidth=0.02,dimen=outer,fillstyle=solid,fillcolor=black](30.126875,3.70375){0.4}
\pscircle[linewidth=0.02,dimen=outer,fillstyle=solid,fillcolor=black](26.126875,-0.29625){0.4}
\pscircle[linewidth=0.02,dimen=outer,fillstyle=solid](22.126875,-0.29625){0.4}
\rput(2.47875,-1.28625){\huge 1}
\rput(26.490938,-1.28625){\huge 7}
\rput(22.503124,-1.28625){\huge 6}
\rput(18.491875,-1.28625){\huge 5}
\rput(14.512813,-1.28625){\huge 4}
\rput(10.4948435,-1.28625){\huge 3}
\rput(6.4889064,-1.28625){\huge 2}
\rput(38.69875,-1.28625){\huge 10}
\pscircle[linewidth=0.02,dimen=outer,fillstyle=solid,fillcolor=black](30.126875,-0.29625){0.4}
\pscircle[linewidth=0.02,dimen=outer,fillstyle=solid](38.126877,-0.29625){0.4}
\pscircle[linewidth=0.02,dimen=outer,fillstyle=solid,,fillcolor=black](34.126877,-0.29625){0.4}
\rput(30.501875,-1.28625){\huge 8}
\rput(34.707813,-1.28625){\huge 9}
\rput(30.702969,2.91375){\huge 11}
\pscircle[linewidth=0.02,dimen=outer,fillstyle=solid](10.126875,-0.29625){0.4}
\pscircle[linewidth=0.02,dimen=outer,fillstyle=solid](6.126875,-0.29625){0.4}
\end{pspicture}
}
\caption{{\protect\footnotesize The $E_{8(-24)}^{+++}$ Tits-Satake diagram.}}
\label{E8-24+++dynkindiagram}
\end{figure}

One can consider as an example the theory based on $\mathbb{O}$, corresponding to the Kac-Moody algebra $E_{8(-24)}^{+++}$, whose Tits-Satake diagram is drawn in fig. \ref{E8-24+++dynkindiagram}. 
Following the standard convention of Tits-Satake diagrams, the black nodes in fig. \ref{E8-24+++dynkindiagram} are associated to compact Cartan generators. By deleting node 3 one obtains the Tits-Satake diagram of $E_{8(-24)}$, which is the global symmetry of the three-dimensional theory, and by further deleting nodes 4, 5 and 6 one obtains the Tits-Satake diagrams of $E_{7(-25)}$, $E_{6(-26)}$ and $SO(1,9)$, which are the global symmetries in four, five and six dimensions, respectively. Node 7, corresponding to a compact Cartan generator, cannot be deleted and the theory cannot be uplifted to seven dimensions. The spectrum of the theory, in the dimensions in which it exists, can be read from Table \ref{E8+++spectrum}, keeping in mind that the reality properties of the various representations differ from those of the maximal theory because the symmetry groups are in different real forms \cite{Riccioni:2008jz}. The same analysis can be performed for the theories based on $\mathbb{H}$, $\mathbb{C}$ and $\mathbb{R}$, whose Tits-Satake very-extended diagrams can be found in  \cite{Riccioni:2008jz}.  It should again be stressed that in all cases the Kac-Moody algebra allows to determine the full
spectrum of the theory, including the $(D-1)$-forms and the $D$-forms.

By analogy with the ${\cal N}=2$ case, it is then natural to associate to the theories based on $\mathbb{C}_{s}$
and $\mathbb{H}_{s}$ the very-extended Kac-Moody algebras $E_{6(6)}^{+++}$
and $E_{7(7)}^{+++}$ \cite{Englert:2003zs,Kleinschmidt:2003mf}; in this
case, the algebra is split and therefore the Tits-Satake diagram coincides
with the Dynkin diagram, with all non-compact Cartan generators. From the
Kac-Moody algebra, one then determines the spectrum of all forms in any
dimension. The highest dimension to which these theories can be uplifted is
10 in the case of $E_{7(7)}^{+++}$ and 8 in the case of $E_{6(6)}^{+++}$.

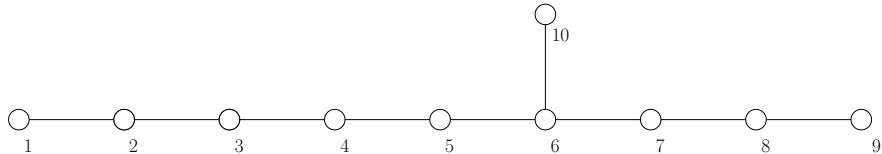
\begin{figure}[h]
\centering
\scalebox{0.35} 
{\
\begin{pspicture}(0,-1.14375)(35.1875,4.10375)
\psline[linewidth=0.02cm](1.926875,-0.29625)(34.126877,-0.29625)
\psline[linewidth=0.02cm](22.126875,3.70375)(22.126875,-0.29625)
\pscircle[linewidth=0.02,dimen=outer,fillstyle=solid](2.126875,-0.29625){0.4}
\pscircle[linewidth=0.02,dimen=outer,fillstyle=solid](6.126875,-0.29625){0.4}
\pscircle[linewidth=0.02,dimen=outer,fillstyle=solid](10.126875,-0.29625){0.4}
\pscircle[linewidth=0.02,dimen=outer,fillstyle=solid](14.126875,-0.29625){0.4}
\pscircle[linewidth=0.02,dimen=outer,fillstyle=solid](18.126875,-0.29625){0.4}
\pscircle[linewidth=0.02,dimen=outer,fillstyle=solid](22.126875,3.70375){0.4}
\pscircle[linewidth=0.02,dimen=outer,fillstyle=solid](26.126875,-0.29625){0.4}
\pscircle[linewidth=0.02,dimen=outer,fillstyle=solid](22.126875,-0.29625){0.4}
\rput(2.47875,-1.28625){\huge 1}
\rput(26.490938,-1.28625){\huge 7}
\rput(22.503124,-1.28625){\huge 6}
\rput(18.491875,-1.28625){\huge 5}
\rput(14.512813,-1.28625){\huge 4}
\rput(10.4948435,-1.28625){\huge 3}
\rput(6.4889064,-1.28625){\huge 2}
\pscircle[linewidth=0.02,dimen=outer,fillstyle=solid](30.126875,-0.29625){0.4}
\pscircle[linewidth=0.02,dimen=outer,fillstyle=solid](34.126877,-0.29625){0.4}
\rput(30.501875,-1.28625){\huge 8}
\rput(34.707813,-1.28625){\huge 9}
\rput(22.702969,2.91375){\huge 10}
\pscircle[linewidth=0.02,dimen=outer,fillstyle=solid](10.126875,-0.29625){0.4}
\pscircle[linewidth=0.02,dimen=outer,fillstyle=solid](6.126875,-0.29625){0.4}
\end{pspicture}
}
\caption{{\protect\footnotesize The $E_{7(7)}^{+++}$ Dynkin diagram.}}
\label{E77+++dynkindiagram}
\end{figure}

We start by considering the $\mathbb{H}_{s}$-based $E_{7(7)}^{+++}$ theory, whose Dynkin diagram is drawn in fig. \ref{E77+++dynkindiagram}.
From the diagram, it is evident that the theory can be uplifted to $D=10$
\cite{Kleinschmidt:2003mf}. Indeed, by deleting node 10, one obtains a
symmetry $SL(10,\mathbb{R})$ associated to a 10-dimensional theory. From the
point of view of the Dynkin diagram, the dimensional reduction corresponds
to further deleting the nodes starting from node 9. The $D$-dimensional
theory corresponds to an $SL(D,\mathbb{R})$ symmetry in the diagram
involving nodes from 1 to $D-1$, and the nodes that are not connected to any
of the $SL(D,\mathbb{R})$ nodes form the global (duality) symmetry 
group. As already mentioned, the deleted node gives an extra scaling symmetry so that the symmetry
associated to the gravity sector is actually $GL(D,\mathbb{R})$, and in the
case in which two nodes have to be cancelled, there is an extra $\mathbb{R}^{+}$ global symmetry corresponding to the fact that there is an additional
internal Cartan generator. Apart from the theories that one obtains from
dimensional reduction of the 10-dimensional theory, there is also an
8-dimensional theory whose corresponding $SL(8,\mathbb{R})$ is formed by the
nodes from 1 to 6 and node 10. We call this theory 8B, whereas we name the
other 8-dimensional theory 8A, in analogy with the maximal case.

\begin{table}[t!]
\renewcommand{\arraystretch}{1.1}
\par
\begin{center}
\scalebox{.9}{
\begin{tabular}{|c|c||c|c|c|c|c|c|c|c|}
\hline
Dim & Symmetry  & $p=1$  & $p=2$ & $p=3$ & $p=4$ & $p=5$ & $p=6$ & $p=7$ & $p=8$  \\
\hline
\hline
10 & $-$ &  &   &    & ${\bf 1}$ &  & &  & \\
\hline 9 & $\mathbb{R}^+$ & ${\bf 1}$ & &${\bf 1}$& ${\bf 1}$ &  &${\bf 1}$ & ${\bf 1}$ &
 \\
 \hline
\multirow{2}{*}{8A} & \multirow{2}{*}{$GL(2,\mathbb{R})$} & \multirow{2}{*}{${\bf 2}$} & \multirow{2}{*}{${\bf 1}$} & \multirow{2}{*}{${\bf 2}$} & \multirow{2}{*}{${\bf 1}$} & \multirow{2}{*}{${\bf 2}$}& ${\bf 3}$& \multirow{2}{*}{$2\times{\bf 2}$}&   ${\bf 3}$\\
& & & & & & & ${\bf 1}$ & & $2\times{\bf 1}$\\
\hline
\multirow{2}{*}{8B }& \multirow{2}{*}{$SL(3,\mathbb{R})$} & & \multirow{2}{*}{${\bf 3}$} &  & \multirow{2}{*}{${\bf \overline{3}}$ }& & \multirow{2}{*}{${\bf 8}$}& & ${\bf 15}$\\
& & & & & & &  & & ${\bf 3}$\\
\hline
& & ${\bf {3}}$ & & & ${\bf \overline{3}}$ &  ${\bf 8}$& ${\bf {8}}$& ${\bf {15}}$ \\
$7$& $GL(3,\mathbb{R})$ & & ${\bf 3}$ &${\bf \overline{3}}$&  &  & ${\bf \overline{6}}$& ${\bf \overline{6}}$ \\
 &  &${\bf 1}$ &  &  & ${\bf 1}$ & ${\bf 1}$& ${\bf 3}$&  $2 \times {\bf 3}$  \\
\cline{1-9}
\multirow{4}{*}{$6$} & \multirow{4}{*}{$SL(4,\mathbb{R})\times SL(2,\mathbb{R})$} &\multirow{4}{*}{${\bf (4,2)}$} & \multirow{4}{*}{${\bf (6,1)}$} & \multirow{4}{*}{${\bf (\overline{4},2)}$} &  & &  ${\bf (64,1)}$\\
& & &  & & ${\bf (15,1)}$ &${\bf (\overline{20},2)}$  & ${\bf (\overline{10},3)}$ \\
& & & & &${\bf (1,3)}$  & ${\bf (4,2)}$ &  ${\bf (6,3)}$ \\
& & & & &  &  &  $2\times {\bf (6,1)}$ \\
\cline{1-8}
& & & & &
${\bf \overline{105}}$  & ${\bf \overline{384}}$  \\
$5$ & $SL(6,\mathbb{R})$ &${\bf 15}$ & ${\bf \overline{15}}$ & ${\bf 35}$ &  & ${\bf 105}$ \\
 & & &  & & ${\bf 21}$ & ${\bf  \overline{15}}$ \\
\cline{1-7}
& & & & & ${\bf 2079}$  \\
$4$ & $SO(6,6)$ &${\bf 32}$ & ${\bf 66}$ & ${\bf 352}$ & ${\bf {462}}$ \\
 & & &  & &  ${\bf 66}$ \\
\cline{1-6}
& & & ${\bf 1539}$& ${\bf 40755}$ \\
$3$ & $E_{7(7)}$ &${\bf 133}$ & &  ${\bf 1539}$ \\
 & & &  ${\bf 1}$ & ${\bf 1}$\\
\cline{1-5}
\end{tabular}
}
\end{center}
\caption{{\protect\footnotesize All the $p$-forms of the $E_{7(7)}^{+++}$
theory in any dimension.}}
\label{E7+++spectrum}
\end{table}

In Table \ref{E7+++spectrum} we list all the $p$-forms of the theory in any
dimension. These (together with gravity and the scalars, that always
parametrise a symmetric manifold $G/H$, where $G$ is the global symmetry and
$H$ is the maximal compact subgroup of $G$) give the full bosonic spectrum
of the theory. In three dimensions, the theory describes 70 scalars, which
is the number of bosonic degrees of freedom in any dimension. In ten
dimensions, the theory contains only a self-dual 4-form \cite%
{Kleinschmidt:2003mf}. By looking at Table \ref{E8+++spectrum}, one can see
that such spectrum arises by truncating the spectrum on the IIB theory to
only singlets of $SL(2,\mathbb{R})$. This fact actually generalises to any
dimension: the symmetry of the maximal theory $G_{D}$ in $D$ dimensions
decomposes as $SL(2,\mathbb{R})\times G_{D}^{(2)}$, where $G_{D}^{(2)}$ is
the $D$-dimensional symmetry of the $E_{7(7)}^{+++}$ theory. Moreover, the
spectrum of the $E_{7(7)}^{+++}$ is obtained by decomposing all
representations of the maximal theory as representations of $SL(2,\mathbb{R})\times G_{D}^{(2)}$ and keeping only those that are singlets of $SL(2,\mathbb{R})$. This truncation is obviously guaranteed to be consistent.

The fact that the $E_{7(7)}^{+++}$ theory is an $SL(2,\mathbb{R})$
truncation of the $E_{8(8)}^{+++}$ theory actually also explains the
occurrence of two different theories (8A and 8B) in eight dimensions.
Indeed, the maximal theory in eight dimensions has symmetry $SL(3,\mathbb{R})\times SL(2,\mathbb{R})$, and thus there are two different ways of
factoring out an $SL(2,\mathbb{R})$. In the first case, taking the $SL(2,\mathbb{R})$ inside $SL(3,\mathbb{R})$, one ends up with the theory denoted
8A in the table, which has symmetry $GL(2,\mathbb{R})$ and arises as the
torus $T^{2}$-reduction of the ten-dimensional theory. On the other hand,
factoring out the other $SL(2,\mathbb{R})$ gives rise to the theory with
global symmetry $SL(3,\mathbb{R})$ denoted 8B in the table, which cannot be
obtained by dimensional reduction.

We can consider in detail how the truncation works in any dimensions by
looking at the representations of the maximal theory that are listed in
Table \ref{E8+++spectrum}. In the case of the ``10B-'' and 9-dimensional case,
the fields in Table \ref{E7+++spectrum} simply correspond to the $SL(2,\mathbb{R})$ singlets in Table \ref{E8+++spectrum}. In eight dimensions, as
mentioned above, the 8A theory corresponds to considering the $SL(2,\mathbb{R})$ inside $SL(3,\mathbb{R})$. The $\mathbf{3}$ and the $\mathbf{\overline{3}}$ both decompose as $\mathbf{2}+\mathbf{1}$, while the $\mathbf{8}$
decomposes as $\mathbf{3}+\mathbf{2}+\mathbf{2}+\mathbf{1}$ and the $\mathbf{6}$ decomposes as $\mathbf{3}+\mathbf{2}+\mathbf{1}$. Taking only the
singlets, this reproduces all the forms up to $p=7$ included in the 8A
theory in Table \ref{E7+++spectrum}. The 8-forms arise from the singlets in
the decompositions of the $\mathbf{15}$ and of the $\mathbf{3}$, which both
have one $SL(2,\mathbb{R})$ singlet. As a result, one
would expect $8$-forms in the $\mathbf{3}$ and $\mathbf{1}$ of the remaining
$SL(2,\mathbb{R})$ as Table \ref{E7+++spectrum} shows, but the singlet appears with multiplicity 2 instead of 3, as the decomposition would suggest.
 The same occurs for the 8B theory, which corresponds to
taking the singlets of the $SL(2,\mathbb{R})$ which is \textit{not} inside $SL(3,\mathbb{R})$ in the eight-dimensional maximal theory. From Table \ref{E8+++spectrum}, one gets that the $8$-forms in the $\mathbf{3}$ should have
multiplicity $2$, while they occur with multiplicity 1 in Table \ref{E7+++spectrum}. 

The reader can check that the same phenomenon occurs in any dimension. All the representations of Table \ref{E7+++spectrum} result from truncating the representations of Table \ref{E8+++spectrum} to the singlets of $SL(2,\mathbb{R})$. Only for the $D$-forms in $D$ dimensions and the 2-forms in three dimensions the multiplicity of the lower-dimensional representations is in general smaller than the multiplicity that results from the decomposition to the singlets of $SL(2,\mathbb{R})$. This can be understood by observing that if a given potential has multiplicity $n$, it means that in the gauge algebra there are actually $n$ potentials  whose gauge 
transformations  with respect to the various gauge parameters of the theory are all different. 
Once all the fields and gauge parameters are truncated to the singlets of $SL(2,\mathbb{R})$, some of the gauge transformations above become identical and the corresponding potentials are identified, with the effect of lowering at least partially the multiplicity.

Given that these theories arise in any dimension as a specific truncation of
the maximal ($E_{8(8)}^{+++}$) theory in which an $SL(2,\mathbb{R})$
subgroup of the global symmetry group is factored out, one can in principle
hope to realise such truncation in perturbative string theory in $D$
dimensions only if this subgroup is part of the perturbative $SO(10-D,10-D)$
symmetry. The highest dimension in which this occurs is $D=8$, where the
perturbative symmetry is the $SL(2,\mathbb{R})\times SL(2,\mathbb{R})$
subgroup of the global symmetry group $SL(3,\mathbb{R})\times SL(2,\mathbb{R})$. In the IIA case, the `geometric' $SL(2,\mathbb{R})$, that is the one
associated to the complex structure of the torus, is the first $SL(2,\mathbb{R})$, namely the one contained in $SL(3,\mathbb{R})$. As a consequence, the
8A theory obtained by factoring out this group could in principle be
obtained from the Type IIA theory \textit{via} some kind of geometric orbifold of the two-torus $T^{2}$. On the other hand, in the IIB case it is the second $SL(2,\mathbb{R})$ which is the `geometric' one. Hence, the 8B theory that
results from factoring out this group could in principle be realised as a geometric orbifold of the Type IIB theory.

The fact that $D=8$ is the highest dimension in which the $\mathbb{H}_{s}$-based
magic non-supersymmetric theory could be realised in perturbative string
theory can also be easily deduced from the $E_{7(7)}^{+++}$ Dynkin diagram in fig. \ref%
{E77+++dynkindiagram}, as already discussed in \cite{Kleinschmidt:2003mf}.
The general rule is that for this to be possible, one must be able to
decompose the global symmetry in $SO(m,m)$ for a given $m$; in the
particular case of the $E_{7(7)}^{+++}$ Dynkin diagram, this decomposition
is achieved by the deletion of node 8. The string dilaton corresponds to the
Cartan generator associated to the simple root $\alpha _{8}$, and the
perturbative symmetry in $D$ dimensions is\footnote{%
This hints the intepretation of $SL(2,\mathbb{R})$ as $Tri(\mathbb{H}_{s})/SO(\mathbb{H}_{s})$, where $Tri(\mathbb{H}_{s})$ and $SO(\mathbb{H}_{s})$ respectively denote the \textit{triality} symmetry and the \textit{%
norm-preserving} symmetry of $\mathbb{H}_{s}$ \cite{triality}.} $SO(8-D,8-D)\times SL(2,\mathbb{R})$, where the second factor corresponds to
the simple root $\alpha _{9}$.

\begin{figure}[h]
\centering
\scalebox{0.35} 
{\
\begin{pspicture}(0,-1.14375)(35.1875,8.10375)
\psline[linewidth=0.02cm](5.926875,-0.29625)(30.126877,-0.29625)
\psline[linewidth=0.02cm](22.126875,7.70375)(22.126875,-0.29625)
\pscircle[linewidth=0.02,dimen=outer,fillstyle=solid](6.126875,-0.29625){0.4}
\pscircle[linewidth=0.02,dimen=outer,fillstyle=solid](10.126875,-0.29625){0.4}
\pscircle[linewidth=0.02,dimen=outer,fillstyle=solid](14.126875,-0.29625){0.4}
\pscircle[linewidth=0.02,dimen=outer,fillstyle=solid](18.126875,-0.29625){0.4}
\pscircle[linewidth=0.02,dimen=outer,fillstyle=solid](22.126875,3.70375){0.4}
\pscircle[linewidth=0.02,dimen=outer,fillstyle=solid](22.126875,7.70375){0.4}
\pscircle[linewidth=0.02,dimen=outer,fillstyle=solid](26.126875,-0.29625){0.4}
\pscircle[linewidth=0.02,dimen=outer,fillstyle=solid](22.126875,-0.29625){0.4}
\rput(26.490938,-1.28625){\huge 6}
\rput(22.503124,-1.28625){\huge 5}
\rput(18.491875,-1.28625){\huge 4}
\rput(14.512813,-1.28625){\huge 3}
\rput(10.4948435,-1.28625){\huge 2}
\rput(6.4889064,-1.28625){\huge 1}
\pscircle[linewidth=0.02,dimen=outer,fillstyle=solid](30.126875,-0.29625){0.4}
\rput(30.501875,-1.28625){\huge 7}
\rput(22.702969,2.91375){\huge 8}
\rput(22.702969,6.91375){\huge 9}
\pscircle[linewidth=0.02,dimen=outer,fillstyle=solid](10.126875,-0.29625){0.4}
\pscircle[linewidth=0.02,dimen=outer,fillstyle=solid](6.126875,-0.29625){0.4}
\end{pspicture}
}
\caption{{\protect\footnotesize The $E_{6(6)}^{+++}$ Dynkin diagram.}}
\label{E66+++dynkindiagram}
\end{figure}
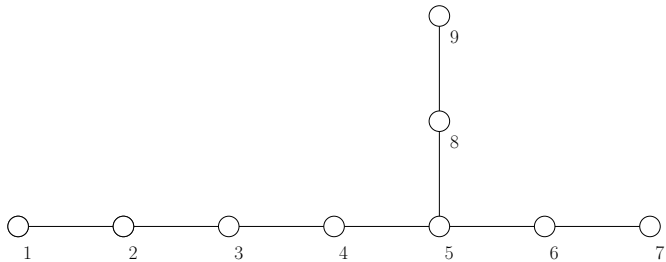

We can repeat the previous analysis for the case of the magic non-supersymmetric theories based on $\mathbb{C}_{s}$, where the symmetry in $D=3$ is $E_{6(6)}$.  In
any dimension, the bosonic sector of the theory can thus be obtained from the Kac-Moody algebra $E_{6(6)}^{+++}$ \cite{Englert:2003zs,Kleinschmidt:2003mf}, whose Dynkin diagram is given in fig. \ref{E66+++dynkindiagram}.
From the diagram, one can deduce the symmetry group in any dimension, together with the highest dimension to which the theory can be uplifted,  $D=8$ in this case. In Table \ref{E6+++spectrum} we give the full spectrum of
$p$-forms of the theory in any allowed dimension $D$, again including, as in the quaternionic case, $D-1$- 
and $D$-forms. The $p$-forms enter the bosonic
sector of the theory together with gravity and with the scalars that
parametrise the manifold $G/H$, being $H$, as usual, the maximal compact
subgroup of the global symmetry group $G$. In particular, in three
dimensions this gives a total of 42 scalars, which is the number of degrees of freedom in any dimension.

Following the same analysis of the previous case, we observe that the
eight-dimensional theory results as a truncation of the maximal theory in
which the $SL(3,\mathbb{R})$ part of the global symmetry is modded out. Once
again, in any dimension the global symmetry $G_{D}^{(3)}$ arises in the
decomposition $G_{D}^{(3)}\times SL(3,\mathbb{R})$ of the symmetry of the
maximal theory, and the full spectrum in dimension $D$ is a consistent truncation of the one
of the maximal theory in dimension $D$, provided only the singlets with respect to the $SL(3,\mathbb{R})$ are kept. The unique exception to this
general rule, already mentioned for the quaternionic theories, occurs for
lower-dimensional representations of the $D$-forms in $D$ dimensions and also of
the 2-forms in $D=3$, whose actual multiplicity is typically lower than the multiplicity resulting from the truncation. As an example, we can analyse 
the 7-dimensional case. By decomposing the $\mathbf{70}$, the $\mathbf{45}$
and the $\mathbf{5}$ of $SL(5,\mathbb{R})$ under the maximal subgroup $GL(2,\mathbb{R})\times SL(3,\mathbb{R})$ and keeping only the $SL(3,\mathbb{R})$-singlets, one gets a $\mathbf{4}$ and three $\mathbf{2}$'s of $SL(2,\mathbb{R})$, while the doublet in Table \ref{E6+++spectrum} only arises with
multiplicity 2.

\begin{table}[t!]
\renewcommand{\arraystretch}{1.1}
\par
\begin{center}
\scalebox{.9}{
\begin{tabular}{|c|c||c|c|c|c|c|c|c|}
\hline
Dim & Symmetry  & $p=1$  & $p=2$ & $p=3$ & $p=4$ & $p=5$ & $p=6$ & $p=7$  \\
\hline
\hline
$8$ & $SL(2,\mathbb{R})$ &  &  & ${\bf 2}$ &  & & ${\bf 3}$&  \\
\hline
\multirow{2}{*}{$7$ }& \multirow{2}{*}{$GL(2,\mathbb{R})$} & \multirow{2}{*}{${\bf 1}$} & \multirow{2}{*}{${\bf 2}$} & \multirow{2}{*}{${\bf {2}}$} & \multirow{2}{*}{${\bf 1}$} &${\bf 3}$ & ${\bf 3}$& ${\bf {4}}$ \\
& & &  & &  &${\bf 1}$  & ${\bf {2}}$&   $2 \times {\bf 2}$  \\
\hline
& & & & &  & ${\bf (3,2)}$   & ${\bf (4,2)}$   \\
& & ${\bf (2,1)}$ & & ${\bf ({2},1)}$ & ${\bf (3,1)}$ & ${\bf (2,3)}$ & ${\bf (2,4)}$  \\
$6$ & $(SL(2,\mathbb{R}))^2\times \mathbb{R}^+$ & & ${\bf (2,2)}$ & & ${\bf (1,1)}$ & &  $3\times{\bf (2,2)}$   \\
& & ${\bf (1,2)}$&  & ${\bf (1,2)}$& ${\bf (1,3)}$ & ${\bf (1,2)}$ & ${\bf (3,1)}$   \\
& & & & &  & ${\bf (2,1)}$ &${\bf (1,3)}$  \\
\cline{1-8}
& & & & & & ${\bf (\overline{15},\overline{3})}$    \\
& & & &${\bf (8,1)}$ & ${\bf (\overline{6},3)}$  & ${\bf (\overline{3},\overline{15})}$   \\
$5$ & $(SL(3,\mathbb{R}))^2$ &${\bf (3,3)}$ & ${\bf (\overline{3},\overline{3})}$ &  & ${\bf (3,3)}$ &$2\times {\bf (\overline{3},\overline{3})}$    \\
 & & &  & ${\bf (1,8)}$ & ${\bf (3,\overline{6})}$ &${\bf (6,\overline{3})}$   \\
& & & & & & ${\bf (\overline{3},6)}$   \\
\cline{1-7}
& & & & ${\bf 70}$ & ${\bf \overline{280}}$  \\
$4$ & $SL(6,\mathbb{R})$ &${\bf 20}$ & ${\bf 35}$ & &  ${\bf {280}}$  \\
 & & &  & ${\bf \overline{70}}$ & ${\bf 189}$   \\

\cline{1-6}
\multirow{4}{*}{$3$} & \multirow{4}{*}{$E_{6(6)}$} &\multirow{4}{*}{${\bf 78}$} &  &  ${\bf \overline{5824}}$ \\
 & & & ${\bf 650}$ & ${\bf 5824}$ \\
& & &${\bf 1}$ & ${\bf 650}$ \\
& & & & ${\bf 78}$ \\
\cline{1-5}
\end{tabular}
}
\end{center}
\caption{{\protect\footnotesize All the $p$-forms of the $E_{6(6)}^{+++}$
theory in any dimension.}}
\label{E6+++spectrum}
\end{table}

One can again determine the dimensions where an interpretation in terms of
a perturbative truncation of the ten-dimensional Type IIA or Type IIB string theory could exist. For this, the requirement that $G\times SL(3,\mathbb{R})$ is contained in the T-duality symmetry of the maximal theory should be
fulfilled. The highest dimension in which this occurs is $D=7$. Again, this
information can be extracted by looking at the Dynkin diagram in fig. \ref%
{E66+++dynkindiagram}. Indeed, an $SO(m,m)$ symmetry only arises after
deleting both nodes 7 and 9, and the dilaton Cartan generator is indeed the
sum of the Cartan generator associated to the simple root $\alpha _{7}$ and
of the one associated to the simple root $\alpha _{9}$. In any dimension,
the perturbative symmetry of the $\mathbb{C}_{s}$-based theory is\footnote{This hints the interpretation of $\mathbb{R}^{+}$ as $Tri(\mathbb{C}_{s})/SO(\mathbb{C}_{s})$, where $Tri(\mathbb{C}_{s})$ and $SO(\mathbb{C}_{s})$
respectively denote the \textit{triality} symmetry and the \textit{norm-preserving} symmetry of $\mathbb{C}_{s}$ \cite{triality}.} $SO(7-D,7-D)\times \mathbb{R}^{+}$.\medskip

To summarise, in this section we have fully characterised the $E_{7(7)}^{+++} $ and $E_{6(6)}^{+++}$ theories, associated respectively to $\mathbb{H}_{s}$ and $\mathbb{C}_{s}$, in terms of consistent $SL(2,\mathbb{R})$ and $SL(3,\mathbb{R})$ truncations of the maximal ($E_{8(8)}^{+++}$)
theory. In the next section we will show that this generalises to further
truncations.

\section{Ehlers $SL(n,\mathbb{R})$ truncations of maximal supergravity in
any dimension}\label{Maximaltrunc}

In the previous section we have shown that the magic non-supersymmetric
theories based on $\mathbb{C}_{s}$ and $\mathbb{H}_{s}$ can be obtained in
any dimensions as suitable truncations of the maximal supergravities. In
particular, the chain of Table \ref{E7+++spectrum} based on $\mathbb{H}_{s}$ is obtained by modding out
the symmetry $SL(2,\mathbb{R})$, while the chain of Table \ref{E6+++spectrum} based on $\mathbb{C}_{s}$
is obtained by modding out the symmetry $SL(3,\mathbb{R})$. In three
dimensions, the modding gives rise to the symmetries $E_{7(7)}$ and $E_{6(6)}$ that coincides with the symmetries of the maximal theory in four and five dimensions,
respectively. The symmetry of the four-dimensional theory based on $\mathbb{C}_{s}$ is $SL(6,\mathbb{R})$, identical to the symmetry of the
five-dimensional theory based on $\mathbb{H}_{s}$. Ultimately, one may trace the previous relations back to the symmetry of the doubly split magic square \cite{Barton:2000ki}.  However, as shown in the previous section, the theories based on $\mathbb{H}_{s}$ and $\mathbb{C}_{s}$ are also associated
to the very-extended Kac-Moody algebras $E_{7(7)}^{+++}$ and $E_{6(6)}^{+++}$.  Moreover, as explained before, the fact that these theories arise as truncations of the maximal
theory can actually be extended to the full spectrum of $p$-forms resulting from the very-extended Kac-Moody algebras.

The aim of this section is to extend this analysis to further truncations.
We consider the truncation of the maximal theory obtained by modding out the U-duality symmetry $G_{D}$ (in any dimension) with respect to the symmetry
$SL(n,\mathbb{R})$. It can be worked out by considering the embedding
\begin{equation}
G_{D}\supset SL(n,\mathbb{R})\times G_{D}^{(n)}\ ,  \label{slntruncation}
\end{equation}
where $G_{D}^{(n)}$ is the residual symmetry of the truncated theory in $D$
dimensions. The resulting groups form the split magic triangle of \cite{Cremmer:1999du} displayed in Table \ref{relevantsubgroups}%
: in its first column, the various $SL(n,\mathbb{R})$'s are reported, while
in the remaining columns the resulting $G_{D}^{(n)}$'s for each $G_{D}$ are
given. In particular, the second column gives the various decompositions of
the three-dimensional theory, the third column gives the various
decompositions of the four-dimensional theory, and so on. Obviously, $G_{D}^{(1)}=G_{D}$ and the $n=2$ and $n=3$ decompositions are just the two
sequences discussed in Section \ref{nonsusytheories}.

\begin{table}[t!]
\renewcommand{\arraystretch}{1.3}
\par
\begin{center}
\scalebox{.55}{
\begin{tabular}{|c||c|c|c|c|c|c|c|c|c|c|}
\hline
$SL(n)$ & $D=3$ & $D=4$ & $D=5$  & $D=6$ & $D=7$ & $D=8$ & $D=9$ & $D=10$A & $D=10$B & $D=11$  \\
\hline
\hline
$n=1$ & $E_{8(8)}$ & $E_{7(7)}$ & $E_{6(6)}$  & $SO(5,5)$ & $SL(5)$ & $SL(3) \times SL(2)$ & $GL(2)$ & $\mathbb{R}^+$ & $SL(2)$ & $1$ \\
\hline $n=2$ & $E_{7(7)}$ & $SO(6,6)$ &$SL(6)$ &$SL(2) \times SL(4)$& $SL(3)\times \mathbb{R}^+$ &$ SL(2 )\times \mathbb{R}^+$ &$\mathbb{R}^+$ & &$1$ &
 \\
& &&&&&$SL(3)$ & & &&\\
\hline
$n=3$&
$ E_{6(6)}$ & $SL(6)$ &$SL(3)\times SL(3)$ &$SL(2) \times SL(2)\times \mathbb{R}^+$& $SL(2)\times \mathbb{R}^+$ &$SL(2 )$ & & & &
 \\
 \hline
$n=4$ & $SO(5,5)$& $SL(4) \times SL(2)$ &  $SL(2) \times SL(2)\times \mathbb{R}^+$ &   $ (\mathbb{R}^+)^2 $ & $\mathbb{R}^+$ & & & & & \\
& & & & $SL(2) \times SL(2)$& & & & & & \\
\hline
$n=5$ & $SL(5)$ &  $SL(3) \times \mathbb{R}^+$&  $SL(2) \times \mathbb{R}^+$ & $ \mathbb{R}^+$ & $1 $ & & & & & \\
\hline
 $n=6$ & $ SL(3) \times SL(2)$ &  $SL(2) \times \mathbb{R}^+$    &  $ SL(2)$  & & & &  & & & \\
 &   & $SL(3)$ & & & & & & & & \\
 \hline
$n=7 $&$SL(2) \times \mathbb{R}^+ $ &  $\mathbb{R}^+$ & & & & & &  & & \\
\hline
$n=8$A &$\mathbb{R}^+$ & & & && & & &&\\
\hline
$n=8$B&$ SL(2)$ & $ 1$ & & & & &  & &&\\
\hline
$n=9$& $1$ & & & & & & & &&\\
\hline

\end{tabular}
}
\end{center}
\caption{{\protect\footnotesize All the relevant regular subgroups that
occur in the $SL(n,\mathbb{R})$ truncations of the U-duality symmetry groups
of maximal supergravity in any dimension, forming the split magic triangle of \cite{Cremmer:1999du}.}}
\label{relevantsubgroups}
\end{table}

For each $n$ (or, equivalently, for each row in the table), the different
entries correspond to the groups that one obtains decomposing the Kac-Moody
algebra $G_{3}^{(n)+++}$ in various dimensions. The $n=2$ and $n=3$ cases
correspond to the theories we discussed in the previous section, while for $n=4$ and $n=5$ one obtains the symmetry groups of the $SO(5,5)^{+++}$ and $SL(5,\mathbb{R})^{+++}$ theories\footnote{%
These are $\mathcal{N}=0$ or $\mathcal{N}=1$ theories in $D=3$ (upliftable
to $\mathcal{N}=0$ theories in $D=4$) \cite{deWit:1992psp, Breitenlohner:1987dg}.}. We will show below that the spectrum of these theories can indeed be
constructed as a consistent truncation of the spectrum of the maximal
theory, in which only singlets of $SL(4,\mathbb{R})$ and $SL(5,\mathbb{R})$
are respectively kept.

If one further extends this identification of the truncated theory with $G_{3}^{(n)+++}$ to higher values of $n$, one gets that the $n=6$ case
corresponds to the $(SL(3,\mathbb{R})\times SL(2,\mathbb{R}))^{+++}$ theory.
In \cite{Kleinschmidt:2008jj}, a way of defining the Kac-Moody very-extended
algebras $G^{+++}$ with $G$ semi-simple but not simple was derived, and a
method to obtain the spectrum of the theory by suitably decomposing the
corresponding Dynkin diagram was given. We will show that, apart from
subtleties concerning the multiplicities of lower-dimensional
representations of the higher-rank forms, the method of \cite%
{Kleinschmidt:2008jj} applied to $(SL(3,\mathbb{R})\times SL(2,\mathbb{R}))^{+++}$ gives results that are in agreement with the $n=6$ truncation.
Going beyond $n=6$, we should point out that there are two possible
decompositions for $n=8$, which we call $n=8$A and $n=8$B. The cases with $n\geq 7$ (with the exception of the $n=8$B case) correspond to very-extended
algebras that are \textit{not} semi-simple. One can easily notice that Table \ref{relevantsubgroups} is \textit{symmetric}, and in particular $G_{D}^{(n)} $ is the same as $G_{n+2}^{(D-2)}$. In other words, the group
that one obtains by modding the $D$-dimensional duality group by $SL(n,\mathbb{R})$ is the same as the one coming out by modding the duality
group in $n+2$ dimensions by $SL(D-2,\mathbb{R})$. This symmetry was firstly explained in \cite{HenryLabordere:2002dk} exploiting the relation discovered in \cite{Iqbal:2001ye} between exceptional groups and del Pezzo surfaces.

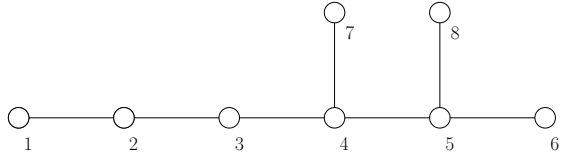
\begin{figure}[h!]
\centering
\scalebox{0.35} 
{\
\begin{pspicture}(0,-1.14375)(32.1875,4.10375)
\psline[linewidth=0.02cm](5.926875,-0.29625)(26.126877,-0.29625)
\psline[linewidth=0.02cm](22.126875,3.70375)(22.126875,-0.29625)
\psline[linewidth=0.02cm](18.126875,3.70375)(18.126875,-0.29625)
\pscircle[linewidth=0.02,dimen=outer,fillstyle=solid](6.126875,-0.29625){0.4}
\pscircle[linewidth=0.02,dimen=outer,fillstyle=solid](10.126875,-0.29625){0.4}
\pscircle[linewidth=0.02,dimen=outer,fillstyle=solid](14.126875,-0.29625){0.4}
\pscircle[linewidth=0.02,dimen=outer,fillstyle=solid](18.126875,-0.29625){0.4}
\pscircle[linewidth=0.02,dimen=outer,fillstyle=solid](22.126875,3.70375){0.4}
\pscircle[linewidth=0.02,dimen=outer,fillstyle=solid](18.126875,3.70375){0.4}
\pscircle[linewidth=0.02,dimen=outer,fillstyle=solid](26.126875,-0.29625){0.4}
\pscircle[linewidth=0.02,dimen=outer,fillstyle=solid](22.126875,-0.29625){0.4}
\rput(26.490938,-1.28625){\huge 6}
\rput(22.503124,-1.28625){\huge 5}
\rput(18.491875,-1.28625){\huge 4}
\rput(14.512813,-1.28625){\huge 3}
\rput(10.4948435,-1.28625){\huge 2}
\rput(6.4889064,-1.28625){\huge 1}
\rput(22.702969,2.91375){\huge 8}
\rput(18.702969,2.91375){\huge 7}
\pscircle[linewidth=0.02,dimen=outer,fillstyle=solid](10.126875,-0.29625){0.4}
\pscircle[linewidth=0.02,dimen=outer,fillstyle=solid](6.126875,-0.29625){0.4}
\end{pspicture}
}
\caption{{\protect\footnotesize The $SO(5,5)^{+++}$ Dynkin diagram.}}
\label{SO55+++dynkindiagram}
\end{figure}

\begin{table}[b!]
\renewcommand{\arraystretch}{1.1}
\par
\begin{center}
\scalebox{.9}{
\begin{tabular}{|c|c||c|c|c|c|c|c|c|}
\hline
Dim & Symmetry  & $p=1$  & $p=2$ & $p=3$ & $p=4$ & $p=5$ & $p=6$ & $p=7$  \\
\hline
\hline
$7$ & $\mathbb{R}^+$ & & ${\bf 1}$ & ${\bf {1}}$ & &${\bf 1}$ & ${\bf 1}$&  ${\bf 1}$  \\
\hline
6A & $(\mathbb{R}^+ )^2$ & $2\times {\bf 1}$&$2\times {\bf 1}$ & $2\times {\bf 1}$ & $2\times {\bf 1}$ &$4\times {\bf 1}$ & $7\times {\bf 1}$    \\
\cline{1-8}
&  & &  &  &  ${\bf (3,1)}$ &  & ${\bf (4,2)}$   \\
6B & $(SL(2,\mathbb{R}) )^2$ & & ${\bf (2,2)}$ &  & & & ${\bf (2,4)}$   \\
&  & &  &  &  ${\bf (1,3)}$ &  & ${\bf (2,2)}$   \\
\cline{1-8}
\multirow{6}{*}{$5$} & \multirow{6}{*}{$(SL(2,\mathbb{R}) )^2 \times \mathbb{R}^+$}& &  &
& & ${\bf (4,2)}$   \\
 & & &  &${\bf (3,1)}$  & ${\bf (3,1)}$  & ${\bf (2,4)}$   \\
& & ${\bf (2,2)}$ &${\bf (2,2)}$  &  \multirow{2}{*}{${\bf (1,3)}$}  &\multirow{2}{*}{${\bf (1,3)}$}  & $3\times {\bf (2,2)}$  \\
& &${\bf (1,1)}$  & ${\bf (1,1)}$ &   & & ${\bf (3,1)}$  \\
& & & & ${\bf (1,1)}$ & $2\times {\bf (2,2)}$&  ${\bf (1,3)}$ \\
& & & &  & &  $2\times {\bf (1,1)}$ \\
\cline{1-7}
& & & & & ${\bf (45,1)}$   \\
& & & ${\bf (15,1)}$ & ${\bf (10,2)}$  & ${\bf (\overline{45},1)}$   \\
$4$ & $SL(4,\mathbb{R})\times SL(2,\mathbb{R})$ &${\bf (6,2)}$ &  &${\bf (\overline{10},2)}$  & $2\times {\bf (15,3)}$    \\
 & & & ${\bf (1,3)}$  & ${\bf (6,2)}$  & $2\times {\bf (15,1)}$ \\
& & & & & ${\bf (1,3)}$   \\
\cline{1-6}
\multirow{6}{*}{$3$} & \multirow{6}{*}{$SO(5,5)$} &\multirow{6}{*}{${\bf 45}$} &  & ${\bf 1050}$   \\
 & & & ${\bf 210}$ & ${\bf \overline{1050}}$  \\
& & & \multirow{2}{*}{${\bf 54}$} & ${\bf 945}$  \\
& & & & ${\bf {210}}$   \\
& & & ${\bf 1}$ & ${\bf {54}}$   \\
& & & & ${\bf {45}}$   \\
\cline{1-5}
\end{tabular}
}
\end{center}
\caption{{\protect\footnotesize All the $p$-forms of the $SO(5,5)^{+++}$
theory in every dimension.}}
\label{SO55+++spectrum}
\end{table}

For the $n=4$ truncation, one obtains the $SO(5,5)^{+++}$ theory, whose
Dynkin diagram is shown in fig. \ref{SO55+++dynkindiagram}. From it, one can
see that the theory can be uplifted at most to seven dimensions, which indeed coincides with the highest dimension in which one can embed $SL(4,\mathbb{R})$ in
the symmetry of the maximal theory. In six dimensions there are two possibilities: the 6A theory, obtained by deleting nodes 6, 7 and 8, is the
dimensional reduction of the seven-dimensional theory, while the 6B theory
corresponds to deleting node 5. The presence of two theories corresponds to the two different embeddings of $SL(4,\mathbb{R})$ inside $SO(5,5)$. In Table \ref{SO55+++spectrum}, we report the spectrum of $p$-forms obtained in various dimensions from the Kac-Moody algebra $SO(5,5)^{+++}$. The representations are those that result from retaining
only the $SL(4,\mathbb{R})$-singlets in the decomposition of the
representations of the symmetry $G_{D}$ of the maximal theory with respect
to $SL(n,\mathbb{R})\times G_{D}^{(n)}$. As in the cases discussed in the
previous section, the only exceptions to this rule are the lower-dimensional
representations of the $D$-forms, which always have actual multiplicity
lower than what would result from the truncation. The number of bosonic
degrees of freedom in any dimension is 25, which is the dimension of the $D=3$ coset manifold $SO(5,5)/[SO(5)\times SO(5)]$.

\begin{table}[t!]
\renewcommand{\arraystretch}{1.1}
\par
\begin{center}
\scalebox{.9}{
\begin{tabular}{|c|c||c|c|c|c|c|c|}
\hline
Dim & Symmetry  & $p=1$  & $p=2$ & $p=3$ & $p=4$ & $p=5$ & $p=6$   \\
\hline
\hline
$7$ & $-$ & &  &  & & &   \\
\hline
$6$ & $\mathbb{R}^+$ & ${\bf 1}$& & $ {\bf 1}$ & ${\bf 1}$ & & $ {\bf 1}$    \\
\cline{1-8}
\multirow{2}{*}{$5$} & \multirow{2}{*}{$GL(2,\mathbb{R})$}& \multirow{2}{*}{${\bf 2}$} &   \multirow{2}{*}{${\bf 2}$} & ${\bf 3}$
&${\bf 2}$ & ${\bf 4}$   \\
 & &  &    & ${\bf 1}$
&${\bf 1}$ & $2\times {\bf 2}$   \\
\cline{1-7}
\multirow{6}{*}{$4$} & \multirow{6}{*}{$GL(3,\mathbb{R})$}& & & & ${\bf 10}$   \\
& & & &${\bf {6}}$ &   ${\bf \overline{10}}$  \\
& &${\bf {3}}$ &${\bf {8}}$ & ${\bf \overline{6}}$ &   $3\times {\bf {8}}$  \\
& & $\overline{\bf {3}}$& ${\bf {1}}$& ${\bf {3}}$&   ${\bf {3}}$  \\
& & & &${\bf \overline{3}}$ &   ${\bf \overline{3}}$  \\
& & & & &   ${\bf {1}}$  \\
\cline{1-6}
\multirow{6}{*}{$3$} & \multirow{6}{*}{$SL(5,\mathbb{R})$} &\multirow{6}{*}{${\bf 24}$} &  & ${\bf 175}$   \\
 & & & ${\bf 75}$ & ${\bf \overline{175}}$  \\
& & & \multirow{2}{*}{${\bf 24}$} & ${\bf 126}$  \\
& & & & ${\bf \overline{126}}$   \\
& & & ${\bf 1}$ & $2\times{\bf {75}}$   \\
& & & & $2 \times {\bf {24}}$   \\
\cline{1-5}
\end{tabular}
}
\end{center}
\caption{{\protect\footnotesize All the $p$-forms of the $SL(5,\mathbb{R}%
)^{+++}$ theory in any dimension.}}
\label{SL5+++spectrum}
\end{table}

In Section \ref{nonsusytheories}, we have discussed the possible realizations within perturbative string theory of the magic non-supersymmetric theories based on $\mathbb{C}_{s}$ and $\mathbb{H}_{s}$. In particular, we have observed that a necessary  condition is that the $SL(2,\mathbb{R})$ or
$SL(3,\mathbb{R})$ symmetry (the factored out term) commutes with the string-dilaton generator. In particular, the theory based on the split quaternions $\mathbb{H}_{s}$ could admit a string interpretation at most in $D=8$, while the theory based on the split complex numbers $\mathbb{C}_{s}$
could be obtained in perturbative string theory at most in $D=7$. In
particular, we showed how this can be read from the Kac-Moody algebra by
looking at the highest dimension in which the symmetry group contains a
subgroup $SO(m,m)$ \cite{Kleinschmidt:2003mf}. In the case of the $%
SO(5,5)^{+++}$ theory, the Dynkin diagram in fig. \ref{SO55+++dynkindiagram}
exhibits a T-duality symmetry already in seven dimensions, corresponding to
the exchange of the nodes 6 and 8. This is in agreement with the fact that $SL(4,\mathbb{R})$ is isomorphic to $SO(3,3)$, and therefore it can be
identified with the perturbative symmetry of the maximal theory in seven
dimensions.

\begin{figure}[h]
\centering
\scalebox{0.35} 
{\
\begin{pspicture}(0,-1.14375)(32.1875,4.10375)
\psline[linewidth=0.02cm](5.926875,-0.29625)(26.126877,-0.29625)
\psline[linewidth=0.02cm](26.126875,3.70375)(26.126875,-0.29625)
\psline[linewidth=0.02cm](14.126875,-0.29625)(26.126875,3.70375)
\pscircle[linewidth=0.02,dimen=outer,fillstyle=solid](6.126875,-0.29625){0.4}
\pscircle[linewidth=0.02,dimen=outer,fillstyle=solid](10.126875,-0.29625){0.4}
\pscircle[linewidth=0.02,dimen=outer,fillstyle=solid](14.126875,-0.29625){0.4}
\pscircle[linewidth=0.02,dimen=outer,fillstyle=solid](18.126875,-0.29625){0.4}
\pscircle[linewidth=0.02,dimen=outer,fillstyle=solid](26.126875,3.70375){0.4}
\pscircle[linewidth=0.02,dimen=outer,fillstyle=solid](26.126875,-0.29625){0.4}
\pscircle[linewidth=0.02,dimen=outer,fillstyle=solid](22.126875,-0.29625){0.4}
\rput(26.490938,-1.28625){\huge 6}
\rput(22.503124,-1.28625){\huge 5}
\rput(18.491875,-1.28625){\huge 4}
\rput(14.512813,-1.28625){\huge 3}
\rput(10.4948435,-1.28625){\huge 2}
\rput(6.4889064,-1.28625){\huge 1}
\rput(26.702969,2.91375){\huge 7}
\pscircle[linewidth=0.02,dimen=outer,fillstyle=solid](10.126875,-0.29625){0.4}
\pscircle[linewidth=0.02,dimen=outer,fillstyle=solid](6.126875,-0.29625){0.4}
\end{pspicture}
}
\caption{{\protect\footnotesize The $SL(5,\mathbb{R})^{+++}$ Dynkin diagram.}
}
\label{SL5+++dynkindiagram}
\end{figure}
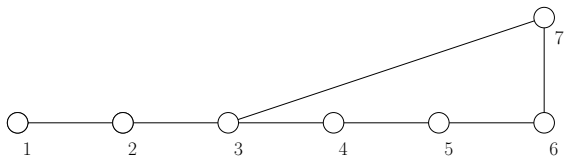

For the $n=5$ truncation, one obtains the $SL(5,\mathbb{R})^{+++}$ theory,
whose Dynkin diagram is shown in fig. \ref{SL5+++dynkindiagram}. 
In this
case, the highest dimension to which this theory can be uplifted is 7,
corresponding to the deletion of node 7 (or node 4). In seven dimensions,
such a theory is nothing but pure gravity \cite{West:2002jj}, in agreement
with the fact that all the fields of the maximal theory in seven dimensions
except the graviton are non-singlets of the global symmetry $SL(5,\mathbb{R}) $. In Table \ref{SL5+++spectrum} we list the spectrum of forms in various
dimensions. The number of bosonic degrees of freedom, coincident with the
dimension of the $D=3$ manifold $SL(5,\mathbb{R})/SO(5)$, is 14.
This theory can admit a possible interpretation in perturbative string
theory starting from five dimensions. Indeed, $SL(5,\mathbb{R})$ can be
embedded in the perturbative symmetry $SO(5,5)$ in five dimensions.
Moreover, looking at the diagram in fig. \ref{SL5+++dynkindiagram}, a
group of $SO(m,m)$ type requires at least the deletion of nodes 5 and
6.

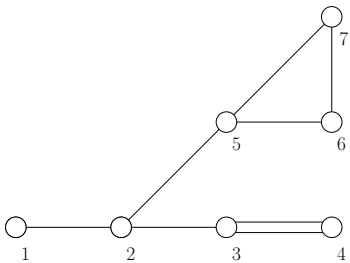
\begin{figure}[h!]
\centering
\scalebox{0.35} 
{\
\begin{pspicture}(0,-1.14375)(24.1875,8.10375)
\psline[linewidth=0.02cm](5.926875,-0.29625)(14.126877,-0.29625)
\psline[linewidth=0.02cm](14.126875,-0.49625)(18.126877,-0.49625)
\psline[linewidth=0.02cm](14.126875,-0.09625)(18.126877,-0.09625)
\psline[linewidth=0.02cm](10.126875,-0.29625)(18.126875,7.70375)

\psline[linewidth=0.02cm](18.126875,7.70375)(18.126875,3.70375)

\psline[linewidth=0.02cm](14.126875,3.70375)(18.126875,3.70375)

\pscircle[linewidth=0.02,dimen=outer,fillstyle=solid](6.126875,-0.29625){0.4}
\pscircle[linewidth=0.02,dimen=outer,fillstyle=solid](10.126875,-0.29625){0.4}
\pscircle[linewidth=0.02,dimen=outer,fillstyle=solid](14.126875,-0.29625){0.4}
\pscircle[linewidth=0.02,dimen=outer,fillstyle=solid](18.126875,-0.29625){0.4}
\pscircle[linewidth=0.02,dimen=outer,fillstyle=solid](14.126875,3.70375){0.4}
\pscircle[linewidth=0.02,dimen=outer,fillstyle=solid](18.126875,3.70375){0.4}
\pscircle[linewidth=0.02,dimen=outer,fillstyle=solid](18.126875,7.70375){0.4}
\rput(18.491875,-1.28625){\huge 4}
\rput(14.512813,-1.28625){\huge 3}
\rput(10.4948435,-1.28625){\huge 2}
\rput(6.4889064,-1.28625){\huge 1}

\rput(18.491875,2.88625){\huge 6}
\rput(14.512813,2.88625){\huge 5}
\rput(18.591875,6.88625){\huge 7}

\pscircle[linewidth=0.02,dimen=outer,fillstyle=solid](10.126875,-0.29625){0.4}
\pscircle[linewidth=0.02,dimen=outer,fillstyle=solid](6.126875,-0.29625){0.4}
\end{pspicture}
}
\caption{{\protect\footnotesize The $( SL(3,\mathbb{R}) \times SL(2,\mathbb{R}) )^{+++}$ Dynkin diagram.}}
\label{SL3SL2+++dynkindiagram}
\end{figure}

Let us now discuss the $n=6$ truncation, corresponding to the $(SL(3,\mathbb{R})\times SL(2,\mathbb{R}))^{+++}$ Kac-Moody algebra\footnote{%
This is an $\mathcal{N}=0$ or $\mathcal{N}=1$ theory in $D=3$ (upliftable to
$\mathcal{N}=0$ in $D=4$) \cite{deWit:1992psp, Breitenlohner:1987dg}.}. It is the
first example, in this context, of a very-extended $G^{+++}$ algebra with $G$
non-simple (namely, semi-simple). As mentioned before, in \cite%
{Kleinschmidt:2008jj} it has been shown that for a Kac-Moody algebra of the
form $(G_{1}\times G_{2})^{+++}$ one can write down a suitable Dynkin
diagram in which the affine nodes of $G_{1}^{+}$ and $G_{2}^{+}$ are
connected. From that diagram, one can then determine the spectrum of the
theory in various dimensions, modulo the subtlety that one has to remove the
extra Cartan generator that \textit{always} occurs in the spectrum. In our
case, the Dynkin diagram is given in fig. \ref{SL3SL2+++dynkindiagram}.
The highest dimension to which the theory can be uplifted is 5, corresponding to
the deletion of nodes 3 and 7, and resulting in the global symmetry $SL(2,\mathbb{R})$. Indeed, $SL(6,\mathbb{R})$ can be embedded in $E_{6(6)}$ but
not in $SO(5,5)$. In four dimensions there are two theories: the 4A is the
reduction of the five-dimensional theory, with a global symmetry $GL(2,\mathbb{R})$ while the 4B has global symmetry $SL(3,\mathbb{R})$. We list  in Table~\ref{SL3SL2+++spectrum} 
the spectrum of forms derived from the Kac-Moody algebra. The number of bosonic degrees of freedom is 7, like the
dimension of the scalar manifold in $D=3$. In any dimension, the spectrum
coincides with truncating the maximal theory to singlets of $SL(6,\mathbb{R}) $, again with the exception of the lower-dimensional representations of
the $D$-forms in $D$ dimensions and the $2$-forms in three dimensions. In
four dimensions, as emerging from the diagram in fig. \ref{SL3SL2+++dynkindiagram} by deleting nodes 4, 6 and 7, one can embed $SL(6,\mathbb{R})$ inside the perturbative symmetry $SO(6,6)$.

\begin{table}[t!]
\renewcommand{\arraystretch}{1.1}
\par
\begin{center}
\scalebox{.9}{
\begin{tabular}{|c|c||c|c|c|c|}
\hline
Dim & Symmetry  & $p=1$  & $p=2$ & $p=3$ & $p=4$    \\
\hline
\hline
{$5$} & {$SL(2,\mathbb{R})$}& &    & ${\bf 3}$
&   \\
\hline
\multirow{2}{*}{4A} & \multirow{2}{*}{$GL(2,\mathbb{R})$}& \multirow{2}{*}{$2\times {\bf 1}$}  & ${\bf 3}$ & \multirow{2}{*}{$2\times  {\bf 3}$}
&   \multirow{2}{*}{$3\times  {\bf 3}$}  \\
& &  & ${\bf 1}$ &
&    \\
\cline{1-6}
& &  &  &
&  ${\bf 10}$   \\
{4B} & {$SL(3,\mathbb{R})$}&  & {${\bf 8}$}  &
& {$  {\bf \overline{10}}$}  \\
& &  &  &
&  ${\bf 8}$   \\
\cline{1-6}
 & & & & ${\bf (10,3)}$ \\
 & & ${\bf (8,1)}$ & ${\bf (8,3)}$ & ${\bf (\overline{10},3)}$ \\
{$3$} & {$SL(3,\mathbb{R})\times SL(2,\mathbb{R})$} & & ${\bf (8,1)}$ & $3\times {\bf (8,3)}$   \\
 & & ${\bf (1,3)}$ & $2\times {\bf (1,1)}$ & $3\times {\bf (8,1)}$
  \\
 & & & & ${\bf (1,3)}$ \\
\cline{1-5}
\end{tabular}
}
\end{center}
\caption{{\protect\footnotesize All the $p$-forms of the $(SL(3,\mathbb{R}%
)\times SL(2,\mathbb{R}))^{+++}$ theory in any dimension.}}
\label{SL3SL2+++spectrum}
\end{table}

\begin{table}[b!]
\renewcommand{\arraystretch}{1.1}
\par
\begin{center}
\scalebox{1}{
\begin{tabular}{|c|c||c|c|}
\hline
Dim & Symmetry  & $p=1$  & $p=2$    \\
\hline
\hline
{$4$} & {$-$}& &       \\
\hline
{$3$} & {$SL(2,\mathbb{R})$}& {$ {\bf 3}$}  & ${\bf 1}$   \\
\hline
\end{tabular}
}
\end{center}
\caption{{\protect\footnotesize All the $p$-forms of the $SL(2,\mathbb{R}%
)^{+++}$ theory in any dimension.}}
\label{SL2+++spectrum}
\end{table}

Among the truncations with $n>6$, the $n=8$B case corresponds to the
Kac-Moody algebra $SL(2,\mathbb{R})^{+++}$, which is pure gravity in four
dimensions \cite{West:2002jj}, $SL(2,\mathbb{R})$ being nothing but the
Ehlers symmetry in the reduction $D=4\rightarrow 3$ of General Relativity
itself \cite{Goroff:1983hc} (also \textit{cfr.} \cite{Ferrara:2012zc}). Indeed,
decomposing the $E_{7(7)}$ representations of the maximal theory in four
dimensions under $SL(8,\mathbb{R})$ one finds that no singlets occur and
therefore only the graviton survives the projection (see Table \ref%
{E8+++spectrum}). The spectrum of forms as derived from the Kac-Moody
algebra is reported in Table \ref{SL2+++spectrum}. In three dimensions the
1-forms are in agreement with the truncation, while there is only one
singlet 2-form instead of the two singlets that would survive the
truncation. There are no 3-forms, while the truncation would give a $\mathbf{3}$ from the $\mathbf{248}$ and a singlet from the $\mathbf{3875}$. For the
highest-dimensional $\mathbf{147250}$ representation of 3-forms in the
maximal theory nothing survives the truncation.

\begin{table}[t!]
\renewcommand{\arraystretch}{1.1}
\par
\begin{center}
\scalebox{.9}{
\begin{tabular}{|c|c|c||c|c|c|c|}
\hline
$n$ & Dim & Symmetry  & $p=1$  & $p=2$ & $p=3$ & $p=4$    \\
\hline
\hline
\multirow{5}{*}{$n=7$} &
{$4$} & {$\mathbb{R}^+$}& & ${\bf 1}$   & $2\times {\bf 1}$
& $2 \times {\bf 1}$  \\
\cline{2-7}
& \multirow{4}{*}{$3$} & \multirow{4}{*}{$GL(2,\mathbb{R})$}& & ${\bf 3}$ &    $2 \times {\bf 4}$  \\
& &  &    ${\bf 3}$ &  \multirow{2}{*}{$2\times {\bf 2}$}
&   $6 \times {\bf 3}$ \\
& &  & ${\bf 1}$ &
&   $6 \times {\bf 2}$ \\
& &  & & $3\times {\bf 1}$
&    $6 \times {\bf 1}$ \\
\cline{1-6}
{$n=8$A}
 & $3$ & $\mathbb{R}^+$ & ${\bf 1}$ & $2\times {\bf 1}$ & $5 \times {\bf 1 }$ \\
\cline{1-6}
$n=9$ & $3$ & $-$ & & ${\bf 1}$ & \\
\cline{1-6}

\end{tabular}
}
\end{center}
\caption{{\protect\footnotesize All the $p$-forms that arise from the $n=7,8$%
A,$9$ truncations of the maximal theory in any dimension.}}
\label{n=78A9truncationspectrum}
\end{table}

The remaining cases $n=7$, 8A and $9$ correspond to \textit{non-semi-simple}
three-dimensional symmetries. The resulting list of the truncations is in
Table \ref{n=78A9truncationspectrum}. The $n=7$ theory exists in four and
three dimensions, while the other two cases only exist in three dimensions.

\section{Ehlers $SL(n,\mathbb{R})$ truncations: non-maximal cases}

\label{nonmaximaltrunc}

In this section we briefly discuss how the analysis carried out in the
previous two sections can be generalised to the 1/2-maximal theories ($16$
supersymmetries) and the 1/4-maximal theories ($8$ supersymmetries). The
crucial difference with respect to the maximal case is that the symmetry groups of these theories are not
in the split real form\footnote{For the relation between these models and del Pezzo surfaces, see \cite{HenryLabordere:2002xh}}. The only exception is the magic theory based on $%
\mathbb{R}$, whose symmetry in three dimensions is $F_{4(4)}$. In \cite{Riccioni:2008jz} it was shown that in general if $G_{3}$ is non-split, then
the very-extended Kac-Moody algebra $G_{3}^{+++}$ corresponding to the
supergravity theory has reality properties that result from considering the
Tits-Satake diagram, which is the one appropriate to identify the real form
of the three-dimensional symmetry $G_{3}$. From it, as in the maximal case,
it is then possible to derive the full spectrum of forms, along with the
surviving real form of the symmetry in any dimensions. Moreover, the nodes
associated to the compact Cartan generators in the Tits-Satake diagram
determine the highest dimension to which the theory can be uplifted.

\begin{table}[t]
\renewcommand{\arraystretch}{1.3}
\par
\begin{center}
\begin{tabular}{|c||c|c|c|c|}
\hline
$SL(n)$ & $D=3$ & $D=4$ & $D=5$ & $D=6$ \\ \hline\hline
$n=1$ & $E_{8(-24)}$ & $E_{7(-25)}$ & $E_{6(-26)}$ & $SO(1,9)$ \\ \hline
$n=2$ & $E_{7(-25)}$ & $SO(2,10)$ &  &  \\ \hline
$n=3$ & $E_{6(-26)}$ &  &  &  \\ \hline
$n=4$ & $SO(1,9)$ &  &  &  \\ \hline\hline
$n=1$ & $E_{7(-5)}$ & $SO^{\ast }(12)$ & $SU^{\ast }(6)$ & $SU(2)\times
SO(1,5)$ \\ \hline
$n=2$ & $SO^{\ast }(12)$ & $SU(2)\times SO(2,6)$ &  &  \\ \hline
$n=3$ & $SU^{\ast }(6)$ &  &  &  \\ \hline
$n=4$ & $SU(2)\times SO(1,5)$ &  &  &  \\ \hline\hline
$n=1$ & $E_{6(2)}$ & $SU(3,3)$ & $SL(3,\mathbb{C})_{\mathbb{R}}$ & $%
U(1)\times SO(1,3) $ \\ \hline
$n=2$ & $SU(3,3)$ & $U(1)\times SO(2,4)$ &  &  \\ \hline
$n=3$ & $SL(3,\mathbb{C})_{\mathbb{R}}$ &  &  &  \\ \hline
$n=4$ & $U(1)\times SO(1,3)$ &  &  &  \\ \hline
\end{tabular}%
\end{center}
\caption{{\protect\footnotesize All the relevant subgroups that occur in the
$SL(n,\mathbb{R})$ truncations of the U-duality symmetry groups for the
1/4-maximal magic theories (also \textit{cfr.} table 2 of \protect\cite{Breitenlohner:1987dg}).}}
\label{relevantsubgroupsN=2magic}
\end{table}

We first consider the 1/4-maximal, magic theories \cite{Gunaydin:1983rk,
Gunaydin:1983bi, Gunaydin:1984ak} based on $\mathbb{C}$, $\mathbb{H}$ and $\mathbb{O}$, and the corresponding decomposition according to equation \eqref{slntruncation} of the U-duality
symmetries. The result is the chain of theories displayed in Table\footnote{Some comments on Table \ref{relevantsubgroupsN=2magic} are in order. The $D=3$ theories with U-duality $E_{6(-26)}$, $SO(1,9)$, $SU^{\ast }(6)$, $SU(2)\times SO(1,5)$, $SL(3,\mathbb{C})_{\mathbb{R}}$ and $U(1)\times
SO(1,3) $ are not present in table 2 of \cite{Breitenlohner:1987dg} because
these theories ($\mathcal{N}=0$ or $\mathcal{N}=1$ in $D=3$) \textit{cannot}
be uplifted to $D=4$. Interestingly, $SO(1,9)$ shares with $F_{4(-20)}$ (the
actual U-duality of the $\mathcal{N}=9$, $D=3$ theory) the same maximal
compact subgroup $SO(9)$, which indeed is the $\mathcal{N}=9$ $\mathcal{R}$-symmetry in $D=3$.  The theory with U-duality $E_{7(-25)}$ in $D=3$ can be $\mathcal{N}=0,1,2$, and in the latter case admits an $\mathcal{N}=1$, $D=4$ uplift to a theory
with U-duality $SO(2,10)$. Analogously, the theory with U-duality $SO^{\ast }(12)$ in $D=3$ can be $\mathcal{N}=0,1,2$, and in the latter case
admits an $\mathcal{N}=1$, $D=4$ uplift to a theory with U-duality $SU(2)\times SO(2,6)$. Moreover, the theory with U-duality $SU(3,3)$ in $D=3 $ can be $\mathcal{N}=0,1,2$, and in the latter case admits an $\mathcal{N}=1 $, $D=4$ uplift to a theory with U-duality $U(1)\times SO(2,4)$.
These three cases can be summarised by stating that the theories with U-duality $Conf\left( \mathbf{J}_{3}^{\mathbb{A}}\right) $ in $D=3$ can be $\mathcal{N}=0,1,2$, and in the latter case they admit an $\mathcal{N}=1$, $D=4$ uplift to a theory with U-duality $(Tri(\mathbb{A})/SO(\mathbb{A}))\times SO(2,q+2)$, where $q:=\dim _{\mathbb{R}}\mathbb{A}=8,4,2$ for $\mathbb{A}=\mathbb{O},\mathbb{H},\mathbb{C}$.} \ref{relevantsubgroupsN=2magic}. As in the analogous Table \ref{relevantsubgroups}
for the maximal theory, Table \ref{relevantsubgroupsN=2magic} is \textit{symmetric}. In all cases, the $n=2$ truncation can be uplifted to four
dimensions, while the $n=3$ and $n=4$ truncations only exist in three
dimensions. In \textit{any} dimension, the spectrum of the theory whose
symmetry is $G_{3}^{(n)}$ in three dimensions can be derived using the
corresponding $G_{3}^{(n)+++}$. The representations of the fields of the
theory based on $\mathbb{O}$, associated to the $E_{8(-24)}^{+++}$ Kac-Moody
algebra, coincide with those of the maximal theory in dimension from three
to six listed in Table \ref{E8+++spectrum}, keeping in mind that the reality
properties of the representations are different because the groups are in
different real forms. Similarly, the spectrum of the $n=2$ truncation of the
theory based on $\mathbb{O}$, associated to the $E_{7(-25)}^{+++}$ Kac-Moody
algebra, can be read by looking at the rows corresponding to $D=3$ and $D=4$
in Table \ref{E7+++spectrum}. This generalises to all the theories listed in
Table \ref{relevantsubgroupsN=2magic}. The analysis of the previous two
sections therefore gives the spectrum of all the theories that are
truncations of the theories based on $\mathbb{O}$, as well as the spectrum
of the theories based on $\mathbb{C}$ and $\mathbb{H}$. The spectrum of the
theories that arise as truncations of the ones based on $\mathbb{C}$ and $\mathbb{H}$ correspond to the Kac-Moody algebras $SO^{\ast }(12)^{+++}$, $SU^{\ast }(6)^{+++}$ and $SU(3,3)^{+++}$, as well as the non-simple cases $(SU(2)\times SO(1,5))^{+++}$ and $SL(3,\mathbb{C})^{+++}$ and the
non-semi-simple case $(U(1)\times SO(1,3))^{+++}$. We have verified that for all the
semi-simple cases the Kac-Moody algebra gives a result consistent with the
truncation in the sense explained in the previous two sections.

The case of the magic theory based on $\mathbb{R}$, corresponding to the
Kac-Moody algebra $F_{4(4)}^{+++}$, is special because the corresponding
symmetry algebra is not simply laced. The symmetry of the theory is $Sp(6,\mathbb{R})$ in $D=4$, $SL(3,\mathbb{R})$ in $D=5$ and $SL(2,\mathbb{R})$ in
$D=6$. The $n=2$ truncation of the three-dimensional theory gives a theory
with symmetry $Sp(6,\mathbb{R})$, whose spectrum can be read from the $Sp(6,\mathbb{R})^{+++}$ Kac-Moody algebra. This theory can be $\mathcal{N}=0,1,2$
in $D=3$, and it can be uplifted at most to $\mathcal{N}=1$, $D=4$, where it
gives a symmetry $SO(2,3)$ which is the $n=2$ truncation of the
four-dimensional $Sp(6,\mathbb{R})$ theory. On the other hand, the theories
in $D=5$ also seem to admit well-defined $n=2$ truncations, but these cannot
result as the uplift of the $Sp(6,\mathbb{R})^{+++}$ because the roots of
the $SL(3,\mathbb{R})$ and the $SL(2,\mathbb{R})$ which are the symmetries
in $D=5$ and $D=6$ are short roots of $F_{4(4)}$. Similarly, the $n=3$ and $n=4$ truncations in three dimensions give theories with symmetries\footnote{%
While the $D=3$ $SL(3,\mathbb{R})$ theory can only be $\mathcal{N}=0,1$ and
it does not admit a supersymmetric uplift to $D=4$, the $D=3$ $SL(2,\mathbb{R})$ theory can be $\mathcal{N}=0,1,2$, and in the latter case it can be
regarded as the dimensional reduction of ``pure'' $\mathcal{N}
=1$, $D=4$ supergravity.} $SL(3,\mathbb{R})$ and the $SL(2,\mathbb{R})$
respectively, but these theories cannot be obtained from the corresponding
very-extended algebras because again the roots of these algebras are short.
What this shows is that in general the truncation analysis is more subtle
for algebras that are not simply laced.

We now move to consider the 1/4-maximal theories whose symmetry is $SO(4,m)$ in three dimensions (related to the cubic semi-simple Jordan
algebra $\mathbb{R}\oplus \mathbf{\Gamma }_{1,m-3}$), as well as to the
1/2-maximal theories whose symmetry is $SO(8,m)$ (related to the cubic
semi-simple Jordan algebra $\mathbb{R}\oplus \mathbf{\Gamma }_{5,m-3}$). We
list the results of the truncations according to equation %
\eqref{slntruncation} in Tables \ref{relevantsubgroupsN=2generic} and \ref{relevantsubgroupsN=4}. As Table \ref{relevantsubgroupsN=2generic} shows, analogously to the magic case, the $n=2$ truncation of the 1/4-maximal theory is
defined in three and four dimensions, while the $n=3$ and $n=4$ truncations
are only defined in three dimensions. There are two different $n=2$
truncations in four dimensions, corresponding to the fact that the $(SL(2,\mathbb{R})\times SO(2,m-2))^{+++}$ Kac-Moody algebra (whose Tits-Satake
diagram can be drawn following the prescription of \cite{Kleinschmidt:2008jj}) admits two different uplifts to four dimensions. Similarly, the
three-dimensional theory admits two different $n=4$ truncations,
corresponding to the fact that there are two theories in six dimensions,
that we call 6A and 6B. Exactly the same considerations apply to the
truncations of the 1/2-maximal theories whose symmetries are listed in
Table \ref{relevantsubgroupsN=4}. In this case
the real form is such that the theory admits an uplift to ten dimensions.

\begin{sidewaystable}
\renewcommand{\arraystretch}{1.6}
\par
\begin{center}
\resizebox{18cm}{!}{
\begin{tabular}{|c||c|c|c|c|c|}
\hline
$SL(n)$ & $D=3$ & $D=4$ & $D=5$ & $D=6$A & $D=6$B\\
\hline
\hline
$n=1$ & $SO(4,m)$ & $SL(2,\mathbb{R}) \times SO(2,m-2)$ & $\mathbb{R}^+ \times SO(1,m-3)$  & $\mathbb{R}^+ \times SO(m-4)$ & $SO(1,m-3)$  \\
\hline $n=2$ & $SL(2,\mathbb{R}) \times SO(2,m-2)$ & $SL(2,\mathbb{R})\times SL(2,\mathbb{R}) \times SO(m-4)$ &  & &
 \\
 & & $SO(2,m-2)$ & & & \\
\hline
$n=3$ & $\mathbb{R}^+ \times SO(1,m-3)$&
 &  & &
 \\
 \hline
$n=4$A & $ \mathbb{R}^+ \times SO(m-4)$& & & & \\
\hline
$n=4$B & $  SO(1,m-3)$& & & & \\
\hline
\end{tabular}
}
\end{center}
\caption{{\footnotesize All the relevant  subgroups of the duality symmetry groups of the  $SO$ theories with 8 supersymmetries.}}
\label{relevantsubgroupsN=2generic}
\vskip 2cm
\renewcommand{\arraystretch}{1.6}
\begin{center}
\resizebox{20cm}{!}{
\begin{tabular}{|c||c|c|c|c|c|c|c|c|c|}
\hline
$SL(n)$ & $D=3$ & $D=4$ & $D=5$  & $D=6$A & $D=6$B & $D=7$ & $D=8$ & $D=9$ & $D=10$  \\
\hline
\hline
$n=1$ & $SO(8,m)$ & $SL(2) \times SO(6,m-2)$ & $\mathbb{R}^+ \times SO(5,m-3)$  & $\mathbb{R}^+ \times SO(4,m-4)$ & $SO(5,m-3)$ & $\mathbb{R}^+ \times SO(3,m-5)$ & $\mathbb{R}^+ \times SO(2,m-6)$ & $\mathbb{R}^+ \times SO(1,m-7)$ & $SO(m-8)$ \\
\hline
$n=2$ & $SL(2) \times SO(6,m-2)$ & $SL(2) \times SL(2) \times SO(4,m-4)$ & $\mathbb{R}^+ \times SL(2) \times SO(3,m-5)$  & $\mathbb{R}^+ \times SL(2) \times SO(2,m-6)$ & $SL(2) \times SO(3,m-5)$ & $\mathbb{R}^+ \times SL(2) \times  SO(1,m-7)$ & $\mathbb{R}^+ \times SL(2) \times SO(m-8)$ &
 &  \\
 &  & $SO(6,m-4)$ &  &  &  &  & &
 &  \\
 \hline
$n=3$ & $\mathbb{R}^+ \times SO(5,m-3)$ & $\mathbb{R}^+ \times SL(2) \times SO(3,m-5)$ & $\mathbb{R}^+ \times \mathbb{R}^+ \times SO(2,m-6)$  & $\mathbb{R}^+ \times \mathbb{R}^+ \times SO(1,m-7)$ & $\mathbb{R}^+ \times SO(2,m-6)$ & $\mathbb{R}^+ \times \mathbb{R}^+ \times  SO(m-8)$ &  &
 &  \\
  \hline
$n=4$A & $\mathbb{R}^+ \times SO(4,m-4)$ & $\mathbb{R}^+ \times SL(2) \times SO(2,m-6)$ & $\mathbb{R}^+ \times \mathbb{R}^+ \times SO(1,m-7)$  & $\mathbb{R}^+ \times \mathbb{R}^+ \times SO(m-8)$ & $\mathbb{R}^+ \times SO(1,m-7)$ &  &  &
 &  \\
   \hline
$n=4$B & $SO(5,m-3)$ & $SL(2) \times SO(3,m-5)$ & $\mathbb{R}^+ \times SO(2,m-6)$  & $ \mathbb{R}^+ \times SO(1,m-7)$ & $SO(2,m-6)$ & $\mathbb{R}^+ \times SO(m-8)$   &  &
 &  \\
   \hline
$n=5$ & $\mathbb{R}^+ \times SO(3,m-5)$ & $\mathbb{R}^+ \times SL(2) \times SO(1,m-7)$ & $\mathbb{R}^+ \times \mathbb{R}^+ \times SO(m-8)$  &  & $\mathbb{R}^+ \times SO(m-8)$ &  &  &
 &  \\
    \hline
$n=6$ & $\mathbb{R}^+ \times SO(2,m-6)$ & $\mathbb{R}^+ \times SL(2) \times SO(m-8)$ & &  &  &  &  &
 &  \\
     \hline
$n=7$ & $\mathbb{R}^+ \times SO(1,m-7)$ &  & &  &  &  &  &
 &  \\
      \hline
$n=8$ & $\mathbb{R}^+ \times SO(m-8)$ &  & &  &  &  &  &
 &  \\
  \hline
\end{tabular}
}
\end{center}
\caption{{\footnotesize All the relevant regular subgroups of the duality symmetry groups in any dimension for theories with 16 supersymmetries.}}
\label{relevantsubgroupsN=4}
\end{sidewaystable}

In general, the embedding that one has to consider for the $SL(n,\mathbb{R})$ truncation of these theories is 
\begin{equation}
SO(q , r  ) \supset SO(n,n) \times SO(q-n, r -n)  \ ,\label{embeddingorthogonal}
\end{equation} 
and therefore the truncation is only possible if both $q$ and $r$ are greater or equal to $n$. For generic $n$, the $SO(n,n)$ subgroup is further decomposed as $\mathbb{R}^+ \times SL(n ,\mathbb{R})$, and the $SL(n,\mathbb{R})$ factor is the one that is truncated. This explains all the entries in Tables \ref{relevantsubgroupsN=2generic} and \ref{relevantsubgroupsN=4}, with the only exceptions of $n=2$ and $n=4$. In the $n=2$ case, for generic $D$, one uses the isomorphism between $SO(2,2)$ and $SL(2,\mathbb{R}) \times SL(2,\mathbb{R})$ and after the truncation an  $SL(2,\mathbb{R})$ subgroup remains.  In $D=4$ the symmetry group is $SL(2,\mathbb{R}) \times SO(2,m-2)$ in the 1/4-maximal case and $SL(2,\mathbb{R}) \times SO(6,m-2)$ in the 1/2-maximal case, and therefore an additional $n=2$ truncation is allowed where the $SL(2,\mathbb{R})$ factor in the symmetry group is truncated out. Finally, for $n=4$, apart from the standard decomposition which is valid for any $n$, one can also consider the embedding in eq. \eqref{embeddingorthogonal} for $n=3$ and identify $SO(3,3)$ with the $SL(4,\mathbb{R})$ that one truncates away. This way of identifying and truncating the $SL(4,\mathbb{R})$ factor gives rise to the $n=4$B theories, while the $n=4$A correspond to the standard identification.

Having explicitly identified the truncation, one can  work out how the various representations of the $p$-form potentials are projected on singlets of $SL(n,\mathbb{R})$. In particular, we
focus on the six-dimensional 1/2-maximal 6A and 6B theories whose symmetry
groups are $\mathbb{R}^{+}\times SO(4,n-4)$ and $SO(5,n-3)$ as reported in
Table \ref{relevantsubgroupsN=4}. The $p$-forms occurring in these theories
are listed in Table \ref{halfmaximalorthogonal1}.
It should be noticed that in the table sets of indices are separated by
commas, where each set corresponds to the antisymmetric indices within a
mixed-symmetry irreducible representation. We want to extract the contributions to the
truncation to singlets of $SL(n,\mathbb{R})$. We can consider for instance  the 4-forms $A_{4,MN}$, where $M,N$ are vector indices of $SO(4,m-4)$ in the 6A theory and of $SO(5,m-3)$ in the 6B theory. The decomposition is 
\begin{equation}
A_{4,MN} \rightarrow A_4 \oplus A_{4,\mu \nu } \ ,
\end{equation}
where the $\mu,\nu$ indices are vector indices of $SO(4-n,m-4-n)$ in 6A and $SO(5-n,m-3-n)$ in 6B.  In this expression, the 4-form singlet that arises is the potential which is dual to the $\mathbb{R}^+$ dilaton that generically occurs in the truncated theory. 
The same decomposition can be worked out for all the representations in Table \ref{halfmaximalorthogonal1},  and as discussed in the previous section one expects that this procedure gives precisely the spectrum of the truncated theory. As in all other cases,  the only exceptions are the 6-forms belonging to the lower-dimensional representations of the symmetry group, whose multiplicity is less than what one would get by truncating on the $SL(n,\mathbb{R})$ singlets.

\begin{table}[t!]
\renewcommand{\arraystretch}{1.4}
\par
\begin{center}
\resizebox{\textwidth}{!}{
\begin{tabular}{|c|c||c|c|c|c|c|c|}
\hline
Dim & Symmetry &$p=1$&$p=2$&$p=3$&$p=4$&$p=5$&$p=6$\\ \hline
6A& $\mathbb{R}^{+}\times SO(4,n-4)$
&
 $A_{1,M}$ &
 $2\times A_2$
&
 $A_{3,M}$
&
 $A_4 \oplus A_{4,MN}$
&
 $2\times A_{5, M}\oplus A_{5,MNP}$
&
 $3\times A_6\oplus 2\times  A_{6,MN}\oplus A_{6, M,N}\oplus A_{6,MNPQ}$ \\ \hline
6B& $SO(5,n-3)$&
&
 $A_{2,M}$
&&
 $A_{4,MN}$
&&
 $A_{6,M}\oplus A_{6,MN,P}$\\ \hline
\end{tabular}
}
\end{center}
\caption{{\protect\footnotesize The $p$-forms in the 6A and 6B theories with
symmetry groups are $\mathbb{R}^{+}\times SO(4,n-4)$ and $SO(5,n-3)$.}}
\label{halfmaximalorthogonal1}
\end{table}

By performing the truncation on all the fields in Table \ref{halfmaximalorthogonal1}, one finds that 
the 6A- and 6B-truncated theories differ with respect
to the initial theories only in the appearance of additional singlets. This
is a completely general result. The $SL(n,\mathbb{R})$ truncation of the
theories with orthogonal symmetry groups $SO(q,r)$ produces a
theory with reduced symmetry $SO(q-n,r-n)$ containing $p$-form potentials that are, rank by
rank, the same tensors of the parent theory, with the only addition of
singlets. The analysis of the four dimensional case can also be similarly carried out but it is slightly more
complicated because of the non-simple symmetry and the fact that the
previous statement holds only in the orthogonal sector. 

\section{Black holes and duality orbits}

\label{BHorbits}

In Section \ref{nonsusytheories} we have analysed the three-dimensional $%
E_{7(7)}$ and $E_{6(6)}$ theories based on split quaternions $\mathbb{H}_{s}$
and split complex numbers $\mathbb{C}_{s}$ respectively, showing how they
can be obtained as truncations of the maximal supergravity and deriving
their uplifts to higher dimensions. In this section, we shall study the
orbit stratification of the relevant representation space of the black-hole
charges under the non-transitive action of the global (duality) symmetry
group. This is particularly relevant in the classification of the (extremal)
black-hole solutions of the corresponding theory. As mentioned above,
together with the three-dimensional maximal theory $E_{8(8)}$ based on split
octonions $\mathbb{O}_{s}$, the two magic non-supersymmetric theories
exhibit, upon dimensional reduction to $D=3$, a duality group of (split)
$E_{n(n)}$ type. The same group can also be realised as \textit{%
quasi-conformal} \cite{Gunaydin:2000xr} group of the corresponding cubic
Jordan algebra $J_{3}^{\mathbb{A}_{s}}$ over the division algebras $\mathbb{A}_{s}=\mathbb{C}_{s},\mathbb{H}_{s},\mathbb{O}_{s}$. The relevant magic
square displaying all the corresponding duality Lie algebras is the
\textit{doubly split} magic square $\mathcal{L}_{3}\left( \mathbb{A}_{s},\mathbb{B}_{s}\right) $ \cite{Barton:2000ki, Gunaydin:2009zza,
Cacciatori:2012cb} given in Table \ref{doublysplitmagicsquare}.

The theory over $\mathbb{O}_{s}$ is maximal supergravity, and the
stratification of U-orbits of asymptotically flat-branes in $D=4,5,6$
dimensions is known \cite{Ferrara:1997uz, Lu:1997bg, Ferrara:1997ci,
Borsten:2010aa, Ferrara:2006xx}. On the other hand, as pointed out above,
the theories over $\mathbb{H}_{s}$ and $\mathbb{C}_{s}$ are \textit{non-supersymmetric}; namely, their bosonic Lagrangian density cannot be
identified with the purely bosonic sector of a supergravity theory. For this
reason, they did not receive great attention in literature\footnote{For symmetries of Freudenthal triple systems and cubic Jordan algebras
defined over split algebras, \textit{cfr. e.g.} \cite{Gunaydin:2000xr}, \cite{Gunaydin:2009zza}, table 1 of \cite{Borsten:2012pd}, and Refs. therein.
Theories over split algebras have been recently considered, in a different
context, in \cite{Bossard:2012ge}. Furthermore, $\mathbb{C}_{s}$- and $\mathbb{H}_{s}$- valued scalar fields have also been recently considered in
cosmology \cite{Gao:2015vca}.}, despite their presence in the classification
of symmetric non-linear sigma models coupled to Maxwell-Einstein gravity (\textit{cfr.} table 2 of \cite{Breitenlohner:1987dg}).

The Jordan algebraic formalism used to classify extremal black hole orbits
in the maximal case in $D=4,5$ can be generalised to the theories based on $\mathbb{C}_{s}$ and $\mathbb{H}_{s}$. In order to show how to proceed, it
can be useful to consider as an example the theories based on $\mathbb{C}_{s} $. In four dimensions, their duality group is $SL(6,\mathbb{R})$ and
the scalar manifold reads
\begin{equation}
\frac{Conf\left( J_{3}^{\mathbb{C}_{s}}\right) }{mcs\left( Conf\left( J_{3}^{\mathbb{C}_{s}}\right) \right) }=\frac{SL(6,\mathbb{R})}{SO(6)}\ ,
\label{Cs-D=4}
\end{equation}
where $Conf\left( J_{3}^{\mathbb{C}_{s}}\right) \simeq Aut\left( \mathfrak{F}\left( J_{3}^{\mathbb{C}_{s}}\right) \right) $ is the \textit{conformal}
group \cite{Gunaydin:2000xr} of the cubic Jordan algebra $J_{3}^{\mathbb{C}_{s}}$ or, equivalently, the automorphism group of the Freudenthal triple
system (FTS) $\mathfrak{F}$ over $J_{3}^{\mathbb{C}_{s}}$ \cite{FTS}, and $mcs$ stands for \textit{maximal compact subgroup}. The $0$-brane (black
hole) dyonic irreducible representation is the rank-3 antisymmetric
self-dual (real) $\mathbf{20}$, so that the pair $\left( SL(6,\mathbb{R}),\mathbf{20}\right) $ defines a group ``of $E_{7}$
-type'', characterised by a unique primitive quartic
invariant polynomial $I_{4}$ \cite{Sato-Kimura, Kac-80, Bermudez:2012fla,
Garibaldi:2013vea}. The action of $SL(6,\mathbb{R})$ on the $\mathbf{20}$
representation determines the stratification into orbits, classified in
terms of invariant constraints on $I_{4}$ or, equivalently, in terms of the
\textit{rank} of the corresponding representative in the Freudenthal triple
system $\mathfrak{F}\left( J_{3}^{\mathbb{C}_{s}}\right) $ \cite%
{rank-J,rank-FTS}. Below we list the stratification together with the
corresponding values of the quartic invariant.

\begin{description}
\item[Rank 1:] The rank 1 orbit is simply
\begin{equation}
\frac{SL(6,\mathbb{R})}{\left[ SL(3,\mathbb{R})\times SL(3,\mathbb{R})\right]
\ltimes \mathbb{R}^{\left( 3,3^{\prime }\right) }}\ .  \label{p-sp-1-II}
\end{equation}

\item[Rank 2:] The rank two orbit reads
\begin{equation}
\frac{SL(6,\mathbb{R})}{\left[ Sp(4,\mathbb{R})\times SO(1,1)\right] \ltimes
\left( \mathbb{R}^{\left( 4,2\right) }\times \mathbb{R}\right) }\ ,
\label{p-sp-2-II}
\end{equation}
where $\mathbb{R}^{\left( 4,2\right) }\simeq \left( \mathbf{4,2}\right) $
denotes the real bi-fundamental\footnote{%
The real fundamental irreducible representation of $Sp(4,\mathbb{R})$ is the
real spinor of $SO(3,2)$.} of the split form $Sp(4,\mathbb{R})\times
SO(1,1)\simeq SO(3,2)\times SO(1,1)$.

\item[Rank 3:] There is only one rank 3 orbit
\begin{equation}
\frac{SL(6,\mathbb{R})}{SL(3,\mathbb{R})\ltimes \mathbb{R}^{8}}\ ,
\label{p-sp-4-II}
\end{equation}%
where $\mathbb{R}^{8}\simeq \mathbf{8}$ denotes the adjoint of $SL(3,\mathbb{%
R})$. 

\item[Rank 4:] In the rank 4 case the quartic invariant $I_{4}$ is different
from zero and there is a splitting of the orbits, depending on the $I_{4}$
sign:
\begin{subequations}
\begin{equation}
I_{4}>0\text{ }:\frac{SL(6,\mathbb{R})}{SL(3,\mathbb{C})_{\mathbb{R}}}
\label{p-sp-5-II}
\end{equation}
and the dyonic
\begin{equation}
I_{4}<0\text{ }:\frac{SL(6,\mathbb{R})}{SL(3,\mathbb{R})\times SL(3,\mathbb{R})}\ ,  \label{p-sp-3-II}
\end{equation}
are the two orbits of rank-4 elements of the FTS $\mathfrak{F}$ over $J_{3}^{\mathbb{C}_{s}}$.
\end{subequations}
\end{description}

It should be noticed that, apart from the rank 4 case where it is induced by
the dyonic solution, for a fixed rank of the FTS there is no stratification
of the orbits. The absence of stratification can be traced back to the
structure of the duality algebra and of its relevant FTS space. In
particular, for maximal theories it has been shown \cite{Ferrara:1997uz,
Lu:1997bg, Ferrara:1997ci, Borsten:2010aa, Ferrara:2006xx} that rank 1
elements in the Jordan algebra construction correspond to single-charge
solutions, while higher rank elements correspond to multi-charge solutions.
Orbits of black hole solution related to different values of the rank can be
computed using bound states of weights in the representation of the duality charges. The number of weights in the bound state must be equal to
the rank in the Jordan algebra construction \cite{Lu:1997bg}. The
stratification reflects the equivalence or the difference between the
considered combinations of weights, while the absence of splitting signals
degeneracy.

\begin{figure}[h]
\centering
\scalebox{0.6} 
{\
\begin{pspicture}(0,-10.347396)(15.865625,10.347396)
\usefont{T1}{ppl}{m}{n}
\rput{35.35122}(-0.5468636,-6.9648657){\rput(10.624531,-4.3246875){$\alpha_{5}$}}
\psline[linewidth=0.02cm,fillcolor=black,dotsize=0.07055555cm 2.0]{*-*}(12.36,-3.3746874)(9.16,-5.5746875)
\psline[linewidth=0.02cm,fillcolor=black,dotsize=0.07055555cm 2.0]{*-*}(4.56,-1.1746875)(7.76,-3.3746874)
\usefont{T1}{ppl}{m}{n}
\rput{-32.069756}(2.0574658,2.957849){\rput(6.1445312,-2.0846875){$\alpha_{1}$}}
\usefont{T1}{ppl}{m}{n}
\rput{35.35122}(1.360323,-5.0696654){\rput(8.604531,-0.3846875){$\alpha_{5}$}}
\psline[linewidth=0.02cm,fillcolor=black,dotsize=0.07055555cm 2.0]{*-*}(10.96,1.0253125)(7.76,-1.1746875)
\psline[linewidth=0.02cm,fillcolor=black,dotsize=0.07055555cm 2.0]{*-*}(7.76,1.0253125)(10.96,-1.1746875)
\usefont{T1}{ppl}{m}{n}
\rput{-32.069756}(1.7499796,5.2879753){\rput(10.044531,-0.3846875){$\alpha_{1}$}}
\psline[linewidth=0.02cm,fillcolor=black,dotsize=0.07055555cm 2.0]{*-*}(7.76,-1.1746875)(7.76,-3.3746874)
\usefont{T1}{ppl}{m}{n}
\rput{90.38534}(5.245192,-10.010972){\rput(7.5645313,-2.3846874){$\alpha_{3}$}}
\usefont{T1}{ppl}{m}{n}
\rput{35.35122}(2.4236658,-3.086462){\rput(6.0245314,2.2753124){$\alpha_{5}$}}
\psline[linewidth=0.02cm,fillcolor=black,dotsize=0.07055555cm 2.0]{*-*}(7.76,3.2253125)(4.56,1.0253125)
\psline[linewidth=0.02cm,fillcolor=black,dotsize=0.07055555cm 2.0]{*-*}(9.16,-5.5746875)(7.76,-7.7746873)
\usefont{T1}{ppl}{m}{n}
\rput{54.790813}(-1.9878318,-9.5652685){\rput(8.204532,-6.6846876){$\alpha_{4}$}}
\psline[linewidth=0.02cm,fillcolor=black,dotsize=0.07055555cm 2.0]{*-*}(4.56,1.0253125)(7.76,-1.1746875)
\usefont{T1}{ppl}{m}{n}
\rput{-32.069756}(0.5884857,3.046651){\rput(5.5645313,0.5153125){$\alpha_{1}$}}
\psline[linewidth=0.02cm,fillcolor=black,dotsize=0.07055555cm 2.0]{*-*}(7.76,-3.3746874)(9.16,-5.5746875)
\usefont{T1}{ppl}{m}{n}
\rput{-58.967655}(7.971037,5.2571054){\rput(8.604531,-4.4046874){$\alpha_{2}$}}
\psline[linewidth=0.02cm,fillcolor=black,dotsize=0.07055555cm 2.0]{*-*}(7.76,-7.7746873)(7.76,-9.974688)
\usefont{T1}{ppl}{m}{n}
\rput{90.38534}(-1.354659,-16.65536){\rput(7.5645313,-8.984688){$\alpha_{3}$}}
\usefont{T1}{ppl}{m}{n}
\rput{35.35122}(3.438426,-1.870806){\rput(4.6245313,4.4753127){$\alpha_{5}$}}
\psline[linewidth=0.02cm,fillcolor=black,dotsize=0.07055555cm 2.0]{*-*}(6.36,5.4253125)(3.16,3.2253125)
\psline[linewidth=0.02cm,fillcolor=black,dotsize=0.07055555cm 2.0]{*-*}(7.76,-3.3746874)(6.36,-5.5746875)
\usefont{T1}{ppl}{m}{n}
\rput{54.790813}(-0.7831279,-7.489834){\rput(6.804531,-4.4846873){$\alpha_{4}$}}
\psline[linewidth=0.02cm,fillcolor=black,dotsize=0.07055555cm 2.0]{*-*}(9.16,5.4253125)(12.36,3.2253125)
\usefont{T1}{ppl}{m}{n}
\rput{-32.069756}(-0.74486357,6.4073696){\rput(10.744532,4.5153127){$\alpha_{1}$}}
\psline[linewidth=0.02cm,fillcolor=black,dotsize=0.07055555cm 2.0]{*-*}(6.36,-5.5746875)(7.76,-7.7746873)
\usefont{T1}{ppl}{m}{n}
\rput{-58.967655}(9.177896,2.9916272){\rput(7.204531,-6.6046877){$\alpha_{2}$}}
\psline[linewidth=0.02cm,fillcolor=black,dotsize=0.07055555cm 2.0]{*-*}(10.96,1.0253125)(10.96,-1.1746875)
\usefont{T1}{ppl}{m}{n}
\rput{90.38534}(10.666663,-10.996103){\rput(10.764531,-0.1846875){$\alpha_{3}$}}
\psline[linewidth=0.02cm,fillcolor=black,dotsize=0.07055555cm 2.0]{*-*}(4.56,-1.1746875)(3.16,-3.3746874)
\usefont{T1}{ppl}{m}{n}
\rput{54.790813}(-0.34060982,-3.9437056){\rput(3.6045313,-2.2846875){$\alpha_{4}$}}
\psline[linewidth=0.02cm,fillcolor=black,dotsize=0.07055555cm 2.0]{*-*}(6.36,5.4253125)(7.76,3.2253125)
\usefont{T1}{ppl}{m}{n}
\rput{-58.967655}(-0.24774492,8.320887){\rput(7.204531,4.3953123){$\alpha_{2}$}}
\psline[linewidth=0.02cm,fillcolor=black,dotsize=0.07055555cm 2.0]{*-*}(4.56,1.0253125)(4.56,-1.1746875)
\usefont{T1}{ppl}{m}{n}
\rput{90.38534}(4.223622,-4.5962486){\rput(4.364531,-0.1846875){$\alpha_{3}$}}
\usefont{T1}{ppl}{m}{n}
\rput{35.35122}(0.8123995,-3.2302678){\rput(5.4445314,-0.3246875){$\alpha_{5}$}}
\psline[linewidth=0.02cm,fillcolor=black,dotsize=0.07055555cm 2.0]{*-*}(7.76,1.0253125)(4.56,-1.1746875)
\psline[linewidth=0.02cm,fillcolor=black,dotsize=0.07055555cm 2.0]{*-*}(12.36,3.2253125)(10.96,1.0253125)
\usefont{T1}{ppl}{m}{n}
\rput{54.790813}(6.5572267,-8.453593){\rput(11.4045315,2.1153126){$\alpha_{4}$}}
\psline[linewidth=0.02cm,fillcolor=black,dotsize=0.07055555cm 2.0]{*-*}(3.16,-3.3746874)(6.36,-5.5746875)
\usefont{T1}{ppl}{m}{n}
\rput{-32.069756}(3.0119221,1.8788022){\rput(4.744531,-4.2846875){$\alpha_{1}$}}
\psline[linewidth=0.02cm,fillcolor=black,dotsize=0.07055555cm 2.0]{*-*}(10.96,-1.1746875)(12.36,-3.3746874)
\usefont{T1}{ppl}{m}{n}
\rput{-58.967655}(7.6362386,9.064961){\rput(11.804531,-2.2046876){$\alpha_{2}$}}
\psline[linewidth=0.02cm,fillcolor=black,dotsize=0.07055555cm 2.0]{*-*}(7.76,3.2253125)(7.76,1.0253125)
\usefont{T1}{ppl}{m}{n}
\rput{90.38534}(9.645093,-5.5813804){\rput(7.5645313,2.0153124){$\alpha_{3}$}}
\psline[linewidth=0.02cm,fillcolor=black,dotsize=0.07055555cm 2.0]{*-*}(9.16,5.4253125)(7.76,3.2253125)
\usefont{T1}{ppl}{m}{n}
\rput{54.790813}(6.9997454,-4.907465){\rput(8.204532,4.3153124){$\alpha_{4}$}}
\psline[linewidth=0.02cm,fillcolor=black,dotsize=0.07055555cm 2.0]{*-*}(3.16,3.2253125)(4.56,1.0253125)
\usefont{T1}{ppl}{m}{n}
\rput{-58.967655}(0.08705309,4.51303){\rput(4.0045314,2.1953125){$\alpha_{2}$}}
\psline[linewidth=0.02cm,fillcolor=black,dotsize=0.07055555cm 2.0]{*-*}(7.76,9.825313)(7.76,7.6253123)
\usefont{T1}{ppl}{m}{n}
\rput{90.38534}(16.244944,1.0630065){\rput(7.5645313,8.615313){$\alpha_{3}$}}
\usefont{T1}{ppl}{m}{n}
\rput{35.35122}(0.46789667,-5.7492094){\rput(9.224531,-2.1246874){$\alpha_{5}$}}
\psline[linewidth=0.02cm,fillcolor=black,dotsize=0.07055555cm 2.0]{*-*}(10.96,-1.1746875)(7.76,-3.3746874)
\psline[linewidth=0.02cm,fillcolor=black,dotsize=0.07055555cm 2.0]{*-*}(7.76,7.6253123)(6.36,5.4253125)
\usefont{T1}{ppl}{m}{n}
\rput{54.790813}(8.204449,-2.8320308){\rput(6.804531,6.5153127){$\alpha_{4}$}}
\psline[linewidth=0.02cm,fillcolor=black,dotsize=0.07055555cm 2.0]{*-*}(7.76,3.2253125)(10.96,1.0253125)
\usefont{T1}{ppl}{m}{n}
\rput{-32.069756}(0.20959252,5.328323){\rput(9.344531,2.3153124){$\alpha_{1}$}}
\psline[linewidth=0.02cm,fillcolor=black,dotsize=0.07055555cm 2.0]{*-*}(7.76,7.6253123)(9.16,5.4253125)
\usefont{T1}{ppl}{m}{n}
\rput{-58.967655}(-1.4546037,10.586365){\rput(8.604531,6.5953126){$\alpha_{2}$}}
\usefont{T1}{ppl}{m}{n}
\rput(7.8009377,9.970312){\Large \psframebox[linewidth=0.02,fillstyle=solid]{0 0 1 0 0}}
\usefont{T1}{ppl}{m}{n}
\rput(7.6909375,7.5703125){\Large \psframebox[linewidth=0.02,fillstyle=solid]{0 1 -1 1 0}}
\usefont{T1}{ppl}{m}{n}
\rput(6.284844,5.3703127){\Large \psframebox[linewidth=0.02,fillstyle=solid]{0 1 0 -1 1}}
\usefont{T1}{ppl}{m}{n}
\rput(3.2848437,3.1703124){\Large \psframebox[linewidth=0.02,fillstyle=solid]{0 1 0 0 -1}}
\usefont{T1}{ppl}{m}{n}
\rput(9.080313,5.3703127){\Large \psframebox[linewidth=0.02,fillstyle=solid]{1 -1 0 1 0}}
\usefont{T1}{ppl}{m}{n}
\rput(7.764219,3.1703124){\Large \psframebox[linewidth=0.02,fillstyle=solid]{1 -1 1 -1 1}}
\usefont{T1}{ppl}{m}{n}
\rput(12.295,3.1703124){\Large \psframebox[linewidth=0.02,fillstyle=solid]{-1 0 0 1 0}}
\usefont{T1}{ppl}{m}{n}
\rput(4.5642185,0.9703125){\Large \psframebox[linewidth=0.02,fillstyle=solid]{1 -1 1 0 -1}}
\usefont{T1}{ppl}{m}{n}
\rput(7.6742187,0.9703125){\Large \psframebox[linewidth=0.02,fillstyle=solid]{1 0 -1 0 1}}
\usefont{T1}{ppl}{m}{n}
\rput(10.578906,0.9703125){\Large \psframebox[linewidth=0.02,fillstyle=solid]{-1 0 1 -1 1}}
\usefont{T1}{ppl}{m}{n}
\rput(4.5642185,-1.2296875){\Large \psframebox[linewidth=0.02,fillstyle=solid]{1 0 -1 1 -1}}
\usefont{T1}{ppl}{m}{n}
\rput(7.7789063,-1.2296875){\Large \psframebox[linewidth=0.02,fillstyle=solid]{-1 0 1 0 -1}}
\usefont{T1}{ppl}{m}{n}
\rput(10.978907,-1.2296875){\Large \psframebox[linewidth=0.02,fillstyle=solid]{-1 1 -1 0 1}}
\usefont{T1}{ppl}{m}{n}
\rput(7.6909375,-10.029688){\Large \psframebox[linewidth=0.02,fillstyle=solid]{0 0 -1 0 0}}
\usefont{T1}{ppl}{m}{n}
\rput(7.7809377,-7.6296873){\Large \psframebox[linewidth=0.02,fillstyle=solid]{0 -1 1 -1 0}}
\usefont{T1}{ppl}{m}{n}
\rput(9.174844,-5.6296873){\Large \psframebox[linewidth=0.02,fillstyle=solid]{0 -1 0 1 -1}}
\usefont{T1}{ppl}{m}{n}
\rput(6.385,-5.6296873){\Large \psframebox[linewidth=0.02,fillstyle=solid]{-1 1 0 -1 0}}
\usefont{T1}{ppl}{m}{n}
\rput(3.2803125,-3.4296875){\Large \psframebox[linewidth=0.02,fillstyle=solid]{1 0 0 -1 0}}
\usefont{T1}{ppl}{m}{n}
\rput(12.284843,-3.4296875){\Large \psframebox[linewidth=0.02,fillstyle=solid]{0 -1 0 0 1}}
\usefont{T1}{ppl}{m}{n}
\rput(7.668906,-3.4296875){\Large \psframebox[linewidth=0.02,fillstyle=solid]{-1 1 -1 1 -1}}
\usefont{T1}{ppl}{m}{n}
\rput(9.422812,9.970312){\Large $\Lambda_{1}$}
\usefont{T1}{ppl}{m}{n}
\rput(4.6228126,5.3703127){\Large $\Sigma_{1}$}
\usefont{T1}{ppl}{m}{n}
\rput(9.422812,7.5703125){\Large $\Lambda_{2}$}
\usefont{T1}{ppl}{m}{n}
\rput(9.622812,3.1703124){\Large $\Lambda_{3}$}
\usefont{T1}{ppl}{m}{n}
\rput(8.422812,1.5703125){\Large $\Lambda_{4}$}
\usefont{T1}{ppl}{m}{n}
\rput(9.622812,-3.4296875){\Large $\Lambda_{6}$}
\usefont{T1}{ppl}{m}{n}
\rput(9.622812,-7.6296873){\Large $\Lambda_{7}$}
\usefont{T1}{ppl}{m}{n}
\rput(7.6228123,-0.4296875){\Large $\Lambda_{5}$}
\usefont{T1}{ppl}{m}{n}
\rput(9.422812,-10.029688){\Large $\Lambda_{8}$}
\usefont{T1}{ppl}{m}{n}
\rput(10.822812,5.3703127){\Large $\Sigma_{2}$}
\usefont{T1}{ppl}{m}{n}
\rput(1.4228125,3.1703124){\Large $\Sigma_{3}$}
\usefont{T1}{ppl}{m}{n}
\rput(2.6228126,0.9703125){\Large $\Sigma_{5}$}
\usefont{T1}{ppl}{m}{n}
\rput(14.022812,3.1703124){\Large $\Sigma_{4}$}
\usefont{T1}{ppl}{m}{n}
\rput(12.422813,0.9703125){\Large $\Sigma_{6}$}
\usefont{T1}{ppl}{m}{n}
\rput(2.6228126,-1.2296875){\Large $\Sigma_{7}$}
\usefont{T1}{ppl}{m}{n}
\rput(1.4228125,-3.4296875){\Large $\Sigma_{9}$}
\usefont{T1}{ppl}{m}{n}
\rput(12.822812,-1.2296875){\Large $\Sigma_{8}$}
\usefont{T1}{ppl}{m}{n}
\rput(14.172813,-3.4296875){\Large $\Sigma_{10}$}
\usefont{T1}{ppl}{m}{n}
\rput(4.5728124,-5.6296873){\Large $\Sigma_{11}$}
\usefont{T1}{ppl}{m}{n}
\rput(11.172813,-5.6296873){\Large $\Sigma_{12}$}
\end{pspicture}
}
\caption{The weights of the $\mathbf{20}$ of $SL(6,\mathbb{R})$.}\label{20sl6dynkintree}
\end{figure}
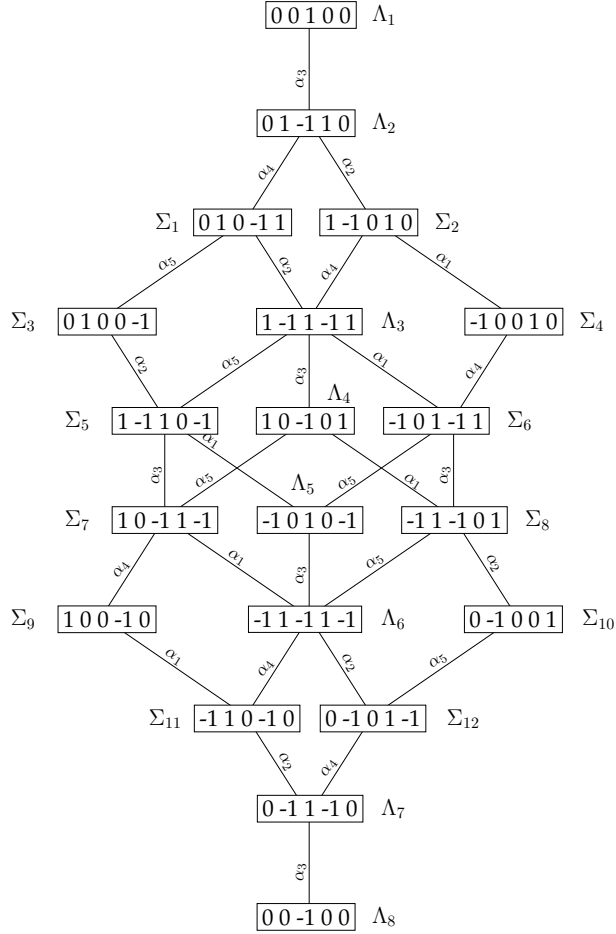

This approach makes it possible to extend this kind of analysis to any
dimension. In order to show how it works, let us analyse the rank 4 orbits
of the previous case; the Dynkin indices of the weights  of the \textbf{20} of $SL(6,\mathbb{R}) $ are shown in fig. \ref{20sl6dynkintree}.
The rank 4 element corresponds to a bound state of the weights $\Lambda
_{1},\ \Lambda _{4},\ \Lambda _{6},\ \Lambda _{7}$. There are three possible
independent bound states that can be written as $\Lambda _{1}+\Lambda
_{4}+a\Lambda _{6}+b\Lambda _{7}$ with $a,b=\pm 1$.\footnote{%
The two states with $a=\pm 1,b=\mp 1$ are not independent.} The stabilizers
are listed in Table \ref{L1L4L6L7stab} as functions of $a$ and $b$.
\begin{table}[h]
\renewcommand{\arraystretch}{1.4}
\par
\begin{center}
\begin{tabular}{|c|c|}
\hline
\textcolor{black}{common} & \multicolumn{1}{|c|}{%
\textcolor{black}{conjunction}} \\ \hline
\textcolor{black}{$\Lambda_1,\Lambda_4, \Lambda_6,\Lambda_7$} & %
\textcolor{black}{$\Lambda_1+\Lambda_4+a \Lambda_6+b\Lambda_7$} \\ \hline
& $E_{\alpha_{2}+\alpha_{3}}-E_{-\alpha_{3}-\alpha_{4}}$ \\
& $E_{\alpha_{3}+\alpha_{4}}-E_{-\alpha_{2}-\alpha_{3}}$ \\
& $E_{\alpha_{1}+\alpha_{2}+\alpha_{3}}-aE_{-\alpha_{3}-\alpha_{4}-%
\alpha_{5}}$ \\
& $E_{\alpha_{3}+\alpha_{4}+\alpha_{5}}-aE_{-\alpha_{1}-\alpha_{2}-%
\alpha_{3}}$ \\
& $E_{\alpha_{5}}-aE_{-\alpha_{1}}$ \\
& $E_{\alpha_{1}}-aE_{-\alpha_{5}}$ \\
& $E_{\alpha_{1}+\alpha_{2}+\alpha_{3}+\alpha_{4}}-bE_{-\alpha_{2}-%
\alpha_{3}-\alpha_{4}-\alpha_{5}}$ \\
& $E_{\alpha_{2}+\alpha_{3}+\alpha_{4}+\alpha_{5}}-bE_{-\alpha_{1}-%
\alpha_{2}-\alpha_{3}-\alpha_{4}}$ \\
& $E_{\alpha_{1}+\alpha_{2}}-bE_{-\alpha_{4}-\alpha_{5}}$ \\
& $E_{\alpha_{4}+\alpha_{5}}-bE_{-\alpha_{1}-\alpha_{2}}$ \\
& $E_{\alpha_{2}}-abE_{-\alpha_{4}}$ \\
& $E_{\alpha_{4}}-abE_{-\alpha_{2}}$ \\
& $F_{\alpha_{3}}^{-ab}+abF_{\alpha_{2}+\alpha_{3}+\alpha_{4}}^{-ab}$ \\
\multirow{-14}{*}{ \begin{tabular}{c} $H_{\alpha_{2}}-H_{\alpha_{4}}$\\
$H_{\alpha_{1}}-H_{\alpha_{5}}$\\ \end{tabular} } & $F_{%
\alpha_{3}}^{-ab}+bF_{\alpha_{1}+\alpha_{2}+\alpha_{3}+\alpha_{4}+%
\alpha_{5}}^{-ab}$ \\ \hline
\end{tabular}
\end{center}
\caption[Stabilizers of $\Lambda _{1}+\Lambda _{4}+a\Lambda _{6}+b\Lambda
_{7}$]{Stabilizers of $\Lambda _{1}+\Lambda _{4}+a\Lambda _{6}+b\Lambda _{7}$%
. The common stabilizers are the generators that annihilate each of the four weights separately, while the conjunction stabilizers are those that give a vanishing result acting on the particular combination of weights considered.}\label{L1L4L6L7stab}
\end{table}
The complexification of the stabilizing algebra gives an $SL(3,%
\mathbb{C})$ with the following generators: 
\begin{align}
H_{\beta _{1}}& =\frac{1}{2}\left[ H_{\alpha _{1}}-H_{\alpha _{5}}+\sqrt{-ab}%
\left( F_{\alpha _{2}+\alpha _{3}+\alpha _{4}}^{-ab}-aF_{\alpha _{1}+\alpha
_{2}+\alpha _{3}+\alpha _{4}+\alpha _{5}}^{-ab}\right) \right] ;  \notag \\
H_{\beta _{2}}& =\frac{1}{2}\left[ H_{\alpha _{2}}-H_{\alpha _{4}}-\sqrt{-ab}%
\left( F_{\alpha _{3}}^{-ab}+abF_{\alpha _{2}+\alpha _{3}+\alpha
_{4}}^{-ab}\right) \right] ,  \notag \\
H_{\beta _{3}}& =\frac{1}{2}\left[ H_{\alpha _{4}}-H_{\alpha _{2}}-\sqrt{-ab}%
\left( F_{\alpha _{3}}^{-ab}+abF_{\alpha _{2}+\alpha _{3}+\alpha
_{4}}^{-ab}\right) \right] ,  \notag \\
H_{\beta _{4}}& =\frac{1}{2}\left[ H_{\alpha _{5}}-H_{\alpha _{1}}+\sqrt{-ab}%
\left( F_{\alpha _{2}+\alpha _{3}+\alpha _{4}}^{-ab}-aF_{\alpha _{1}+\alpha
_{2}+\alpha _{3}+\alpha _{4}+\alpha _{5}}^{-ab}\right) \right] ,
\end{align}%
\begin{align}
E_{\beta _{1}}& =E_{\alpha _{1}}-aE_{-\alpha _{5}}-\sqrt{-ab}\left(
E_{\alpha _{1}+\alpha _{2}+\alpha _{3}+\alpha _{4}}-bE_{-\alpha _{2}-\alpha
_{3}-\alpha _{4}-\alpha _{5}}\right) ,  \notag \\
E_{\beta _{2}}& =E_{\alpha _{2}+\alpha _{3}}-E_{-\alpha _{3}-\alpha _{4}}+%
\sqrt{-ab}\left( E_{\alpha _{2}}-abE_{-\alpha _{4}}\right) ,  \notag \\
E_{\beta _{3}}& =E_{\alpha _{3}+\alpha _{4}}-E_{-\alpha _{2}-\alpha _{3}}+%
\sqrt{-ab}\left( E_{\alpha _{4}}-abE_{-\alpha _{2}}\right) ,  \notag \\
E_{\beta _{4}}& =E_{\alpha _{5}}-aE_{-\alpha _{1}}-\sqrt{-ab}\left(
E_{\alpha _{2}+\alpha _{3}+\alpha _{4}+\alpha _{5}}-bE_{-\alpha _{1}-\alpha
_{2}-\alpha _{3}-\alpha _{4}}\right) ,
\end{align}
\begin{align}
E_{-\beta _{1}}& =E_{\alpha _{5}}-aE_{-\alpha _{1}}+\sqrt{-ab}\left(
E_{\alpha _{2}+\alpha _{3}+\alpha _{4}+\alpha _{5}}-bE_{-\alpha _{1}-\alpha
_{2}-\alpha _{3}-\alpha _{4}}\right) ,  \notag \\
E_{-\beta _{2}}& =E_{\alpha _{3}+\alpha _{4}}-E_{-\alpha _{2}-\alpha _{3}}-%
\sqrt{-ab}\left( E_{\alpha _{4}}-abE_{-\alpha _{2}}\right) ,  \notag \\
E_{-\beta _{3}}& =E_{\alpha _{2}+\alpha _{3}}-E_{-\alpha _{3}-\alpha _{4}}-%
\sqrt{-ab}\left( E_{\alpha _{2}}-abE_{-\alpha _{4}}\right) ,  \notag \\
E_{-\beta _{4}}& =E_{\alpha _{1}}-aE_{-\alpha _{5}}+\sqrt{-ab}\left(
E_{\alpha _{1}+\alpha _{2}+\alpha _{3}+\alpha _{4}}-bE_{-\alpha _{2}-\alpha
_{3}-\alpha _{4}-\alpha _{5}}\right) ,
\end{align}
\begin{align}
E_{\beta _{1}+\beta _{2}}& =E_{\alpha _{1}+\alpha _{2}+\alpha
_{3}}-aE_{-\alpha _{3}-\alpha _{4}-\alpha _{5}}-\sqrt{-ab}\left( E_{\alpha
_{1}+\alpha _{2}}-bE_{-\alpha _{4}-\alpha _{5}}\right) ,  \notag \\
E_{\beta _{3}+\beta _{4}}& =E_{\alpha _{3}+\alpha _{4}+\alpha
_{5}}-aE_{-\alpha _{1}-\alpha _{2}-\alpha _{3}}-\sqrt{-ab}\left( E_{\alpha
_{4}+\alpha _{5}}-bE_{-\alpha _{1}-\alpha _{2}}\right) ,  \notag \\
E_{-\beta _{1}-\beta _{2}}& =E_{\alpha _{3}+\alpha _{4}+\alpha
_{5}}-aE_{-\alpha _{1}-\alpha _{2}-\alpha _{3}}+\sqrt{-ab}\left( E_{\alpha
_{4}+\alpha _{5}}-bE_{-\alpha _{1}-\alpha _{2}}\right) ,  \notag \\
E_{-\beta _{3}-\beta _{4}}& =E_{\alpha _{1}+\alpha _{2}+\alpha
_{3}}-aE_{-\alpha _{3}-\alpha _{4}-\alpha _{5}}+\sqrt{-ab}\left( E_{\alpha
_{1}+\alpha _{2}}-bE_{-\alpha _{4}-\alpha _{5}}\right) .
\end{align}
By varying the values of $a$ and $b$, one obtains two different real forms
for the stabilizers, namely $SL(3,\mathbb{C})_{\mathbb{R}}$ and $%
SL(3,\mathbb{R})\times SL(3,\mathbb{R})$,
corresponding respectively to $a=b=\pm 1$ and $a=-b=\pm 1$ choices\footnote{The subscript ``$\mathbb{R}$'' denotes the Lie algebra to be
considered as an algebra over the reals.}. 
The two resulting real forms, $SL(3,\mathbb{C})_{\mathbb{R}}$ and
$SL(3,\mathbb{R})\times SL(3,\mathbb{R})$ of $SL(3,\mathbb{C})$, have the same signature, but they are
discriminated by looking at the imaginary units appearing in the Chevalley
basis (in particular, for $SL(3,\mathbb{C})_{\mathbb{R}}$ there
are not imaginary units in the stabilizing algebra). Summarizing, the
4-weights bound state orbits are those given in \eqref{p-sp-5-II} and \eqref{p-sp-3-II}. 
It should be stressed that, although in principle the independent bound
states would have been three, only two orbits are present. One is the dyonic
orbit, corresponding to $a$ and $b$ with opposite signs. The second orbit is
related to the two combinations where $a$ and $b$ have the same sign. They
are independent, but give rise to the \textit{same} orbit, explaining the
\textit{absence} of stratification. The same behaviour is exhibited by the
theories based on $\mathbb{H}_{s}$, extending this property to the whole
family of theories based on $\mathbb{C}_{s}$, $\mathbb{H}_{s}$ and $\mathbb{O%
}_{s}$.

It is interesting to compare the previous set of theories to magic $\mathcal{%
N}=2$ and $\mathcal{N}=4$ supergravity theories in $D=4$ where, on the
contrary, a rich stratification of orbits appears (for a comprehensive
treatment, see \cite{Borsten:2011ai}). To understand the basic factors marking
such a difference it is worth to consider, as a representative example, the $\mathcal{N}=2$ magic supergravity based on $\mathbf{J}_{3}^{\mathbb{R}}$,
whose $D=4$ U-duality group is $Sp(6,\mathbb{R})$, obtained uplifting the $F_{4(4)}$ three-dimensional theory. In this case, the $0$-branes (black
holes) belong to the $\mathbf{{14^{\prime }}}$ (rank-$3$ antisymmetric
skew-traceless) irreducible representation, whose highest weight, $\Lambda _{1}$, is $\boxed{1\
0\ 0}$. To construct the rank two orbit, we have to combine the highest
weight with the weight $\Lambda _{4}$ identified by the Dynkin labels $\boxed{2\ 0\ -1}$. The possible independent bound states are thus $\Lambda
_{1}\pm \Lambda _{4}$. The rank two-orbits read
\begin{flalign}
 &\frac{Sp(6,\mathbb{R})}{SO(1,3)\ltimes \mathbb{R}^{4}\times\mathbb{R}}
\end{flalign}and
\begin{flalign}
 &\frac{Sp(6,\mathbb{R})}{SO(2,2)\ltimes \mathbb{R}^{4}\times\mathbb{R}} \ ,
\end{flalign}corresponding to the combinations with the plus and minus sign,
respectively. Thus, two non-isomorphic rank-2 orbits exist. The same
splitting phenomenon takes place for the rank-three and rank-four cases, namely two rank-three orbits and three rank-four orbits are present. It is
worth remarking that the maximal possible splitting is actually realised,
since two and three are respectively the number of independent three- and
four- charge bound states. 
The obtained stratification of orbits is not surprising, being related to
the presence of weights of different lengths in the $\mathbf{{14^{\prime }}}$, a property never encountered in the supergravity theories related to split
composition algebras \cite{Marrani:2015gfa}. Indeed, the short weights are
responsible for the change of compactness of some generators in the
stabilizer when switching from one combination to the other, giving rise to
the split of the orbits. In particular, in the case at hand, the conjunction
stabilizer $F_{\alpha _{2}+\alpha _{3}}^{\pm }=E_{\alpha _{2}+\alpha
_{3}}\pm E_{-\alpha _{2}-\alpha _{3}}$ appearing in the full set of
stabilizer of Table \ref{stabilizers2-ll14primesp6} does the job.
\begin{table}[h]
\renewcommand{\arraystretch}{1}
\par
\begin{center}
\begin{tabular}{|c|c|c|}
\hline
common & \multicolumn{2}{|c|}{conjunction} \\ \hline
$\Lambda_{1},\Lambda_{4}$ & $\Lambda_{1}+\Lambda_{4}$ & $\Lambda_{1}-%
\Lambda_{4}$ \\ \hline\hline
$E_{2\alpha_{1}+2\alpha_{2}+\alpha_{3}}$ &  &  \\
$E_{\alpha_{1}+2\alpha_{2}+\alpha_{3}}$ & \multirow{-2}{*}{$F_{\alpha_{2}+%
\alpha_{3}}^{-}$} & \multirow{-2}{*}{$F_{\alpha_{2}+\alpha_{3}}^{+}$} \\
$E_{\alpha_{1}+\alpha_{2}+\alpha_{3}}$ &  &  \\
$E_{\alpha_{1}+\alpha_{2}}$ & \multirow{-2}{*}{$E_{2\alpha_{2}+%
\alpha_{3}}-E_{-\alpha_{3}}$} & \multirow{-2}{*}{$E_{2\alpha_{2}+%
\alpha_{3}}+E_{-\alpha_{3}}$} \\
$E_{\alpha_{1}}\quad E_{\alpha_{2}}$ &  &  \\
$H_{\alpha_{2}}$ & \multirow{-2}{*}{$E_{\alpha_{3}}-E_{-2\alpha_{2}-%
\alpha_{3}}$} & \multirow{-2}{*}{$E_{\alpha_{3}}+E_{-2\alpha_{2}-%
\alpha_{3}}$} \\
$E_{-\alpha_{2}}$ &  &  \\ \hline
\end{tabular}%
\end{center}
\caption{Stabilizers for the 2-weights bound states $\Lambda _{1}\pm \Lambda
_{4}$ in the $\mathbf{14}^{\prime }$ of $Sp(6,\mathbb{R})$.}
\label{stabilizers2-ll14primesp6}
\end{table}

The previous considerations hold true not only for the uplifts of the $F_{4(4)}$ theory but also for its $SL(n,\mathbb{R})$ truncations. This is
the case, for instance, of the three dimensional theory\footnote{%
Note that $F_{4(4)}$ embeds non-symmetrically and maximally two $SL(3,\mathbb{R})$'s, which are not on the same footing. The one yielding triplets
and anti-triplets in the decomposition of the adjoint of $F_{4(4)}$ is the
Ehlers group.} with $SL(3,\mathbb{R})$ obtained by truncating the Ehlers $SL(3,\mathbb{R})$. In this theory the 0-branes belong to the representation 
\footnote{This is nothing but the representation of $\mathbf{J}_{3}^{\mathbb{R}}$ with
respect to its reduced structure group $SL(3,\mathbb{R})$. Indeed, $SL(3,\mathbb{R})$ is also the global (U-duality) symmetry of the $\mathcal{N}=2$, $D=5$ uplift of the $\mathcal{N}=4$, $D=3$ $F_{4(4)}$ theory.} $\mathbf{6}$
that, containing again weights of two different lengths, induces the
splitting of the orbits. The phenomenon is completely general, and an
exhaustive treatment will be presented in a forthcoming paper \cite{progress}.

An analysis of the other $\mathcal{N}=2$ ($\mathbb{R}\oplus \mathbf{\Gamma }_{1,m-3}$-based) and the $\mathcal{N}=4$ ($\mathbb{R}\oplus \mathbf{\Gamma }_{5,m-3}$-based) supergravity theories points out additional subtleties. In
particular, since the U-duality symmetries do not occur in the split form,
non-real weights are present in their representations. The reality
properties of the weights depend on the real form of the algebra, and they
are encoded in the corresponding Tits-Satake diagram. Non-real weights play
here the same role as the short weights, giving rise in an analogous way to
orbit splittings \cite{Marrani:2015gfa}. Let us explain the mechanism using
as a guide the mentioned analysis of the rank-four orbits in the $\mathbf{20}$ of $SL(6,\mathbb{R})$. If the algebra had been $SU(3,3)$ instead of $SL(6,\mathbb{R})$, the degeneracy between the
two rank four orbits with $a=b=\pm 1 $ would have been lifted, leaving three
distinct rank four orbits, namely $SL(6,\mathbb{R})/[SU(3)\times SU(3)]$ for
$a=b=1$, $SL(6,\mathbb{R})/[SU(1,2)\times SU(1,2)]$ for $a=b=-1$ and $SL(6,\mathbb{R})/[SL(3,\mathbb{C})_{\mathbb{R}}]$ in the other cases (note that
the first two cosets do not exist!).

The picture emerging from the previous discussion seems to point towards a
precise statement: in absence of supersymmetry there is not splitting of
orbits, while in non-maximal supergravity theories the orbit splitting can
take place depending on the real form and on the relevant representations of
the duality group. 

\section{Discussion and conclusions}

\label{conclusions}

We have analysed the magic non-supersymmetric theories based on split quaternions $\mathbb{H}_{s}$ and split
complex numbers $\mathbb{C}_{s}$. These theories can
be obtained as  $SL(2,\mathbb{R})$ and $SL(3,\mathbb{R})$ Ehlers truncations of maximal supergravity and are related to the $E_{7(7)}^{+++}$ and the $E_{6(6)}^{+++}$ very
extended Kac-Moody algebras \cite{Englert:2003zs,Kleinschmidt:2003mf}. 
We have generalised the procedure to $SL(n,\mathbb{R})$ Ehlers truncations (with $n>3$) of the maximal supergravity giving rise to additional classes of non-supersymetric theories, as well as to $SL(n,\mathbb{R})$ Ehlers
truncations of non-maximal supergravity theories. It should be emphasised
that our analysis involves not only the propagating degrees of freedom but
also the $(D-1)$- and $D$-forms, in any dimension $D\geq 3$. 
Since the field strength of the $(D-1)$-forms are dual to mass parameters, our analysis also encodes massive deformation and gaugings.  In some cases, the
truncation generates theories that are obtained by very-extended Kac-Moody algebras of the  form $[G_{1}\times G_2 ]^{+++}$, that  have been introduced in \cite{Kleinschmidt:2008jj}. Finally, we have discussed  properties of the duality orbits 
of extremal black-hole solutions of these theories.

An interesting issue is related to the embedding of the class of magic
theories into perturbative string theory\footnote{See also Sections 12 and 13 of \cite{Gunaydin:2009pk}.}. In \cite{Bianchi:2007va}, the magic exceptional
supergravity based on $\mathbf{J}_{3}^{\mathbb{O}}$ and with an $E_{8(-24)}$
symmetry in $D=3$ has been constructed in $3\leq D \leq 6$. In particular, the
six-dimensional theory has been identified as the long wavelength limit of a
certain compactification of Type IIB on $K3$. It is realised as a
peculiar shift-orientifold\cite{Angelantonj:2002ct,Rohm:1983aq,Kounnas:1988ye,Kiritsis:1997ca,Antoniadis:1998ki,Antoniadis:1998ep,Vafa:1995gm} of the Type IIB, where the unoriented projection
truncates the $5$ tensor multiplets of the untwisted sector to $1$ tensor
multiplet and $4$ hypermultiplets and the $16$ twisted tensor multiplets to $8$ tensor multiplets and $8$ hypermultiplets. The introduction of $16$
D5-branes to cancel anomalies provides $16$ additional abelian vector
multiplets, with gauge group $U(1)^{16}$. The momentum shift paired to the $\mathbb{Z}_{2}$-orbifold involution defining the $K3$  prevents the introduction of D9-branes.
The other $D$-dimensional models ($D<6$) in the chain are obtained by reducing the
six-dimensional theory on a $(6-D)$-torus. In particular, the magic octonionic
four-dimensional theory with $E_{7(-25)}$ symmetry is thus obtained
as a freely-acting orientifold of Type IIB on $K3\times T^{2}$.

There also exist string-theory realizations of the complex and quaternionic magic theories in four dimensions. In \cite{Sen:1995ff},  the theory defined by the algebra $J_{3}^{\mathbb{H}}$ has been built as an asymmetric shift-orbifold of the Type IIA string.  In particular, starting from an (S-)dual pair of Type IIA theories compactified on $T^4$ in six dimensions and performing a suitable asymmetric shift-orbifold projection on $T^2$ \cite{Rohm:1983aq,Kounnas:1988ye,Kiritsis:1997ca,Vafa:1995gm}, one ends up with a {\it self dual} theory in four dimensions, exactly coincident with the magic theory.  Interestingly, the theory is free of hypermultiplets and with the dilaton belonging to a vector multiplet.  The {\it bosonic}  massless spectrum includes the $\mathcal{N}=2$ gravity multiplet coupled to 15 vector multiplets and perfectly  coincides with the one of the $\mathcal{N}=6$ pure supergravity, obtainable with analogous construction in terms of another self-dual theory.
The $30$ scalars, of course, parametrise the coset $SO^*(12)/U(6)$.

The same quaternionic magic model has been 
obtained in \cite{Dolivet:2007sz} as $\mathcal{N}=2$ (non-geometric) compactifications of Type IIA, using a different asymmetric shift-orbifolds realised within the free fermionic construction\cite{Antoniadis:1986rn, Kawai:1986ah}. The procedure rests on adding a peculiar chiral twist that substitutes the extra gravitinos with fermions in the twin $\mathcal{N}=6$ model, related to a simpler asymmetric shift projection.  Again, these models are free of hypermultiplets with the dilaton belonging to a vector multiplet.
Using an analogous procedure, the magic theory defined by $J_{3}^{\mathbb{C}}$ can be obtained as a projection from the $\mathcal{N}=3$ theory coupled to 3 vector multiplets, realised again as an asymmetric shift-orbifold of the Type IIA with free fermions. The massless spectrum contains an $\mathcal{N}=2$ supergravity coupled to 8 vector multiplets. The 18 scalars parametrise the coset $SU(3,3)/SU(3)\times SU(3) \times U(1)$, the model is hyper-free and the dilaton is in a vector multiplet.  It should be stressed that these compactifications have $(1,4)$ supersymmetry on the world-sheet and do not correspond to Calabi-Yau compactifications, associated to $(2,2)$ supersymmetry on the world-sheet.   The quaternionic model can be uplifted to five
dimensions and reduced on $S^{1}$ to three dimensions, while the complex
model can be reduced to three dimensions but cannot be oxidised to higher dimensions because the
involved moduli come from twisted sectors of the orbifold.  As emerging from the previous discussion, it is clear that 
the embedding within string theory is not necessarily unique. For instance, besides the two realizations of the magic quaternionic theory in four dimensions, the
five-dimensional magic quaternionic theory can also be obtained \cite{Bianchi:2007va} as an $S^{1}$ compactification of a six dimensional orientifold of a corresponding Gepner model \cite{Angelantonj:1996mw}.

The situation is subtler for the magic non-supersymmetric theories, where
quantum corrections are not protected. As seen, two are the
``necessary'' conditions: one is that the
dilaton must factor out of the truncation algebra, the second is that the
truncation algebra itself must be a sub-algebra of the perturbative
T-duality symmetry. In general, it is not obvious that even if the two
conditions are respected, the model can be seen as a perturbative truncation
of a certain string model whose massless sector coincides with the
non-truncated supergravity. Just to give a taste of the problem, let us
consider one of the simplest models, the 8B theory in eight dimensions
related to split quaternions. If realised in string theory, it should result
as a truncation to eight dimensions of the ten dimensional Type IIB string,
whose massless bosonic NS-NS sector coincides with the one of the IIB
supergravity. Specifically, it contains the graviton, $g_{MN}$, the two form
$B_{MN}$ and the dilaton $\varphi $ in the NS-NS sector and a scalar $C_{0}$%
, a two-form $C_{MN}$ and a self-dual four form $C_{MNPQ}^{+}$ in the R-R
sector. By compactifying on a two-torus down to eight dimensions, the
spectrum can be organised in terms of representations of the geometric $SL(2)$ group. The truncated theory corresponds to keeping only the $SL(2)$
singlets. It amounts to have a non-supersymmetric theory with a massless
spectrum consisting of the graviton $g_{\mu \nu }$, the dilaton $\varphi $,
four additional scalars $\varphi _{i},i=1,...4$ and three two-forms. The
scalars correspond to the internal part of the ten-dimensional two-forms $B$
and $C$, to the volume of the two torus and to the surviving R-R scalar. The
two forms are the survival spacetime components of the ten dimensional $B$
and $C$ and an additional two form coming from the combination of the
internal components of the self-dual ten dimensional four-form. The question
is whether the described remnant spectrum in eight dimensions is obtainable
as a string theory projection of the Type IIB. In other words, we need a
compactification on a manifold that is able to throw away the fermions and
to project the rest on singlets of $SL(2,\mathbb{Z})$. The simplest natural
action one could envisage, as in the magic supersymmetric cases, is a freely
acting (Scherk-Schwarz) orbifold deformation (like that in \cite{Rohm:1983aq,Kounnas:1988ye,Kiritsis:1997ca,Antoniadis:1998ki,Antoniadis:1998ep,Vafa:1995gm})
combined with the action of a discrete (finite) subgroup commuting with, or
stabilizing, the $SL(2,\mathbb{Z})$. The most promising attempt, a freely
acting $\mathbb{Z}_{4}$ orbifold, does the truncation job but unfortunately
does not exist at the level of perturbative string theory, since a modular
invariant $\mathbb{Z}_{4}$ orbifold projection of Type IIB is not available
in eight dimensions. We cannot suggest closer models nor give definite
answers. As said, since we are singling out electric-magnetic duality symmetries, it is
not at all guaranteed that a perturbative string theory model exists
corresponding to these truncations. It could be that some reductions of this
theory exist in lower dimensions but there are obstructions to oxidise them
up to eight dimensions. Of course, it would also be very interesting to
analyse non-perturbative completions of our models but, being non
supersymmetric, the control over quantum corrections is unavoidably very
limited.

\vskip 1cm

\section*{Acknowledgements}

L.R. would like to thank E. Bergshoeff and the Quantum Gravity group of the Rijksuniversiteit Groningen for their kind hospitality  while this work was being completed.  This work has been supported in part by the STaI ``Uncovering Excellence''
Grant of the University of Rome ``Tor Vergata'', CUP
E82L15000300005 and in part by ``Sapienza'' University of Rome.

\end{document}